\documentclass[a4paper,11pt]{article}
\pdfoutput=1 

\usepackage{jheppub} 
\usepackage[T1]{fontenc} 
\usepackage{float}
\usepackage{color}

\title{\boldmath Information recovery in the Hayden-Preskill protocol}

\author[a]{Bart{\l}omiej Czech,}
\author[a]{Sirui Shuai,}
\author[b,a]{and Haifeng Tang}
\affiliation[a]{Institute for Advanced Study, Tsinghua University, Beijing 100084, China}
\affiliation[b]{382 Via Pueblo Mall, Stanford University, Stanford CA 94305, USA}

\emailAdd{bartlomiej.czech@gmail.com}
\emailAdd{siruishuai@gmail.com}
\emailAdd{danniiiocean@gmail.com}

\abstract{We revisit information retrieval from evaporating black holes in the Hayden-Preskill protocol, treating the black hole dynamics as Haar-random. We compute, down to the first exponentially suppressed terms, all integer-indexed R{\'e}nyi mutual informations between a black hole, its radiation, and a reference that catalogues Alice's diaries. We find that dropping a diary into a young black hole effectively delays the Page time. We also compute the radiation : diary reflected R{\'e}nyi entropies, and identify a technical reason why they cannot be continued to the reflected entropy by the replica trick.}

\begin{document} 
\maketitle
\flushbottom

\section{Introduction}

In their seminal work~\cite{Hayden2007mirror}, Hayden and Preskill argued that black holes can act as information mirrors. That is, contrary to common sense and Hawking's calculation~\cite{Bekenstein1973BHentropy, Hawking1973FourLaws, Bekenstein1974generalized, Hawking1975particle}, information dropped into a black hole not only returns to the outside world but moreover, in some cases, does so instantaneously. Importantly, black holes attain this mirroring property with age: old black holes reflect infalling information right away whereas young black holes withhold the information until they become old.

The distinction between young and old black holes has recently taken on a central significance. The cross-over time when a black hole officially becomes old---the Page time~\cite{Page1993average, Page1993information, Page2013time}---is when the growth in the black hole's radiation entropy turns around. It is also the time when the black hole interior becomes an island~\cite{Penington2020entanglementWedge,Almheiri2019entropy,Almheiri2020page}---that is when, according to the rules of holographic subregion duality~\cite{Czech2012gravity, Dong:2016eik}, it becomes reconstructible from previously expelled Hawking radiation. In the modern view, the transition from the no-island to the yes-island regime is understood as a phase transition in the erasure correcting capacity of the bulk spacetime~\cite{RTfromError}: an island is a region that can be recovered from the Hawking radiation even if we `erase' the entire exterior spacetime.

\begin{figure}[t]
\centering
\includegraphics[width=0.8\textwidth]{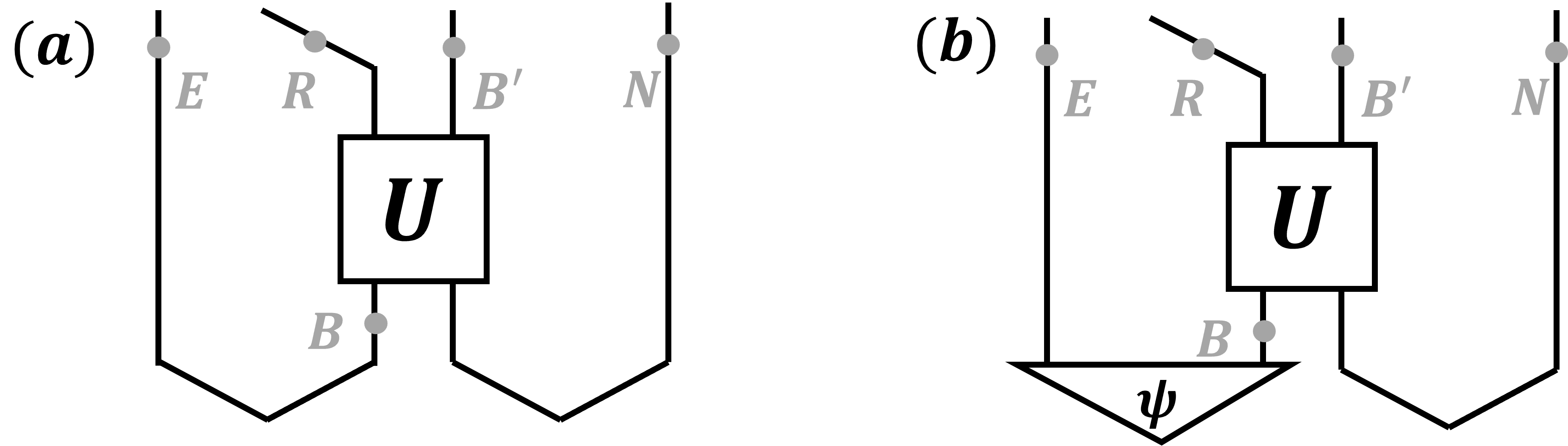}
\label{fig:HP-protocol}
\caption{The setup of the Hayden-Preskill protocol for an old (a) and young (b) black hole.}
\end{figure}

This recent progress makes it pertinent to reexamine the Hayden-Preskill conclusion in quantitative ways. This is the subject of the present paper. To fix the narrative---and as a brief review---we consider the following setup:
\begin{itemize}
\item Alice drops diary $A$ into a black hole $B$. To keep track of different versions of Alice's diaries, we let $A$ be maximally entangled with a diary-reference $N$. Questions about recovery of Alice's diary will be phrased in terms of mutual informations $I(\ldots:N)$ involving the system $N$.
\item The initial black hole is denoted $B$. The quantum states of system $B$ are the microstates of the black hole. The age of the black hole is modeled by the degree of entanglement between $B$ and its previously expelled radiation $E$:
\begin{itemize}
\item For old black holes, $E$ and $B$ are maximally entangled.
\item For young black holes, $E$ and $B$ are assumed to form some random pure state $|\psi\rangle$. The assumption of randomness is motivated either by the scrambling property of black holes~\cite{susskind2008fast} or simply by the fact that we do not know the dynamics of quantum gravity.
\end{itemize}
\item After Alice drops her diary, we model the subsequent dynamics as a Haar-random unitary $U: BA \to B'R$. Since diary $A$ is maximally entangled with reference $N$, we equivalently study unitaries~~$U: BN \to B'R$. Here $R$ denotes the subsequent Hawking radiation (after Alice's diary was minced by the black hole) and $B'$ is the state of the black hole after $R$ was radiated away. The assumed randomness of $U$ is justified the same way as the randomness of $|\psi\rangle$ above.
\item The quantity that controls whether Alice's diary can be recovered from Bob by collecting both early radiation $E$ and subsequent radiation $R$ is the mutual information $I(ER:N)$.
\end{itemize}
The setup is illustrated in Figure~\ref{fig:HP-protocol}. 

Hayden and Preskill did not directly compute $I(ER:N)$ in their original paper. Instead, they proved that when the black hole is old, $ER$ contains a subsystem $\hat{M}$ such that the reduced state $\rho_{\hat{M}N}$ has nearly maximal fidelity with the maximally entangled pure state on $\hat{M}N$. This ensures that recovery of information from $\hat{M} \subset ER$ is possible. However, their calculation is rather indirect and leaves out certain interesting aspects of the protocol, which can otherwise be computed and characterized with ease. An example question of this type is: How does Alice's tossing of the diary affect the Page time---understood as a transition between the no-recovery and yes-recovery eras? (We answer that question at the end of Section~\ref{sec:young}.)

We have not found in the literature an analytic characterization of $I(ER:N)$ or the state of radiation $\rho_{ER}$. One exception is the series of works~\cite{Yoshida2017efficient, Yoshida2019soft, Yoshida2022recovery}, where the R{\'e}nyi entropy
\begin{equation}
S^{(q)}(\rho) = (1-q)^{-1} \log {\rm tr} \,\rho^q
\end{equation}
and the R{\'e}nyi entropy mutual information
\begin{equation}
I^{(q)}(ER:N) = S^{(q)}(\rho_{ER}) + S^{(q)}(\rho_{N}) - S^{(q)}(\rho_{ERN})
\end{equation}
was computed for $q=2$. Numerically, $I(ER:N)$ and $\rho_{ER}$ were explored in ~\cite{Bae:2019niw}. Other recent relevant work on the Hayden-Preskill protocol includes \cite{Lie:2022xjb, Tajima:2021nwu, Bao:2020zdo, Nakata:2020vvy} in quantum information theory, \cite{Garcia2022abt, Leone:2022afi, Vardhan:2021mdy} in holographic duality, and \cite{Nakata2023HPinHamitonian, Hayata:2021kcp, Blok:2020may, Cheng:2019yib} in quantum many-body physics. None of them, however, compute the higher-$q$ R{\'e}nyi entropies and mutual informations. This is the primary technical deliverable in the present paper.

We compute the R{\'e}nyi entropies $S^{(q)}(\rho)$ for arbitrary integer $q$, in old and young black holes, for the radiation ($ER$), diary reference ($N$) and the remaining black hole ($B'$) systems, down to the first exponentially suppressed order. These quantities are easily converted into R{\'e}nyi mutual informations, spectral densities (see Appendix~\ref{sec:spectra}) and---by the replica trick---into ordinary mutual informations $I(X:Y)$. An interesting outcome of our results is a characterization of a transitional range of radiation, at which the trio diary-radiation-black hole all have nonvanishing pairwise mutual informations. We highlight this finding in Figure~\ref{fig:entropies} and elaborate on it in the Discussion.

In addition, we also attempt a replica trick computation of the reflected mutual information $I_R(ER:N)$, as defined in \cite{dutta2021canonical}. This quantity has recently attracted interest in holographic, condensed matter, and information theoretic contexts. However, the reflected R{\'e}nyi entropies we find do not admit an analytic continuation to the reflected von Neumann entropy {\`a} la replica trick. We discuss the technical reason for this, which is that the loop counting which undergirds the calculation differs qualitatively between replica indices $q = 1$ and $q \geq 2$. To the extent that the reflected R{\'e}nyi entropies we compute reliably reflect the reflected entanglement spectrum, we find it to be flatter than the non-reflected spectrum of the young black hole.

All our calculations depend crucially on the assumed random character of the time evolution $U$ and the black hole-radiation state $|\psi\rangle$. The same methodology---based on integrals over the unitary group called Weingarten functions---has recently been used in a variety of contexts, see for example~\cite{XLQ2020random, nonisometriccodes, subleadingweingartens, Vijay2023random}. Of course, the exact quantum gravity time evolution operator might not be usefully approximated as random. In that case, our results serve as a benchmark to quantify this non-randomness.

\paragraph{Notation} To ease the notation, we will use the uppercase label of each subsystem to also denote the dimension of its Hilbert space, for example $\dim \mathcal{H}_N := N$. To express entropies, we will often `count degrees of freedom' in base $e$. When doing so, we will use the same letter as the system label but in lowercase, e.g. $N = \exp(n)$. Readers who prefer to think in qubits will want to divide by $\log 2$; for example, $n = \log N$ means that system $N$ can be said to contain $n/\log 2$ qubits. 

\paragraph{Organization} In Section~\ref{sec:renyi} we compute R{\'e}nyi entropies of the radiation $ER$ and the black hole $B'$ (the purifier of the combined system $ERN$), and consequently their R{\'e}nyi mutual informations. Section~\ref{sec:reflected} calculates reflected R{\'e}nyi entropies and discusses the obstacles, which prevent an analytic continuation to the reflected entropy. We close with a Discussion and two appendices. Appendix~\ref{sec:weingarten} reviews Weingarten functions while Appendix~\ref{sec:spectra} converts the R{\'e}nyi entropies to entanglement spectra. 

\section{R{\'e}nyi mutual information in the Hayden-Preskill protocol}
\label{sec:renyi}
The setup of the calculation is reviewed in and below Figure~\ref{fig:HP-protocol}. Our basic objective is to compute the $I^{(q)}(ER:N)$ for integer $q$, averaged over the unitary $U$ with uniform measure. We distinguish two cases: old black holes and young black holes. In the latter case, we also average over the initial state $|\psi\rangle$ of $EB$---the initial black hole-radiation system. 

Our calculation assumes that Alice's diary is much smaller than the black hole: $N \ll B$. On the logarithmic scale (counting degrees of freedom using $\log B := b$ etc.), our assumption means:
\begin{equation}
b \gg n + c = r
\end{equation}
Here $c$ quantifies the extra overhead in radiation, which Hayden and Preskill concluded is necessary to decode Alice's diary. That is, if the diary contains $n / \log 2$ qubits, Bob will attempt to decode it after capturing $(n+c)/\log 2$ qubits of extra radiation. 

The calculations in this section confirm but also extend the conclusions in \cite{Hayden2007mirror}. The main new results in Section~\ref{sec:old} are the $U$-averaged R{\'e}nyi entropies of the combined radiation system $ER$. They are given in equations (\ref{iqlargee}), (\ref{iqlargeenfinal}) and (\ref{iqlargeenc}). The main novelty in Section~\ref{sec:young} is an analysis of the range of radiation, in which the mutual information $I(ER:N)$ transitions from approximately 0 to $2n$ (double diary size). It is given in equation~(\ref{rrange}) and illustrated in Figure~\ref{fig:entropies}. We also inspect in detail the regime where the collected radiation is comparable to the `Page gap'---the gap that separates the black hole from reaching Page time and becoming old (Section~\ref{sec:midage}).

\subsection{Old black hole}
\label{sec:old}
We first notice that $\rho_N$ and $\rho_{B'}$ (which is purified by $ERN$) are maximally mixed in their respective Hilbert spaces. This is readily recognized from the diagrammatic expressions for the density operators, from which $U$ and $U^\dagger$ cancel out.\footnote{In Appendix~\ref{sec:weingarten} we compute ${\rm tr}(\rho_{B'})^q = B'^{1-q}$ without using the cancelation of $U$ and $U^\dagger$. That calculation serves as a reference for several other computations in this paper, and showcases a useful property of integrals over the unitary group.} Our task therefore reduces to computing $S^{(q)}(\rho_{ER})$.

\paragraph{Structure of calculation} 
The trace of $(\rho_{ER})^q$ is a periodic array of the following form:
\begin{equation}
\centering
\includegraphics[width=0.8\textwidth]{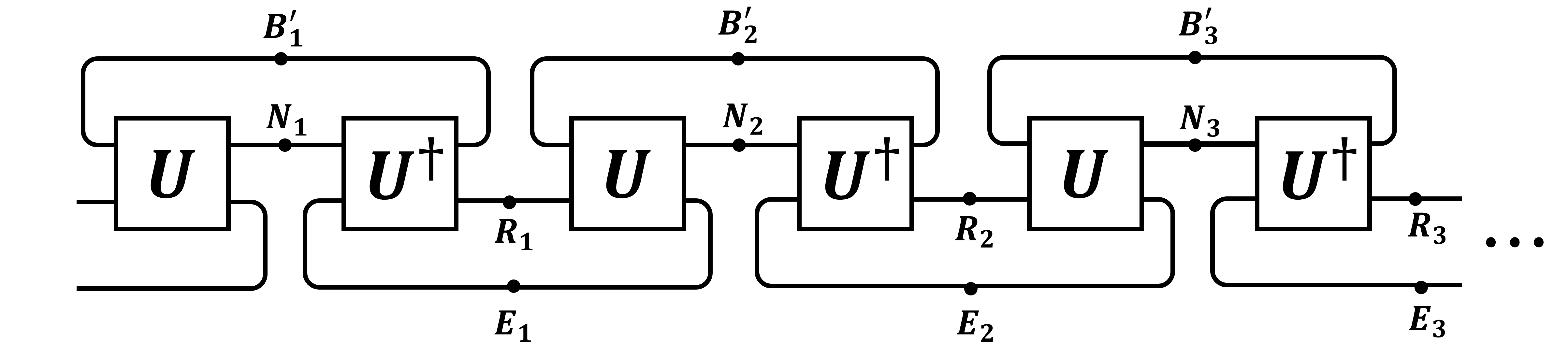}
\label{eq:oldstructure}
\end{equation}
Our task is to average this expression over $U \in U(EN)$. For this purpose, we need the average matrix element of $q$ copies of $U$'s and $U^\dagger$s. This matrix element has one collective index in $q$ copies of $EN$ and one collective index in $q$ copies of $RB'$. We will label the $q$ copies of each subsystem in subscript. As an example, the triplicated system $(EN)^{\otimes 3}$ has a basis of the form:
\begin{equation}
| i \rangle_{(EN)_1} \otimes | j \rangle_{(EN)_2} \otimes | k \rangle_{(EN)_3}
\end{equation}

The requisite average matrix element is given in terms of Weingarten functions; see Appendix~\ref{sec:weingarten}:
\begin{equation}
\centering
\includegraphics[width=0.8\textwidth]{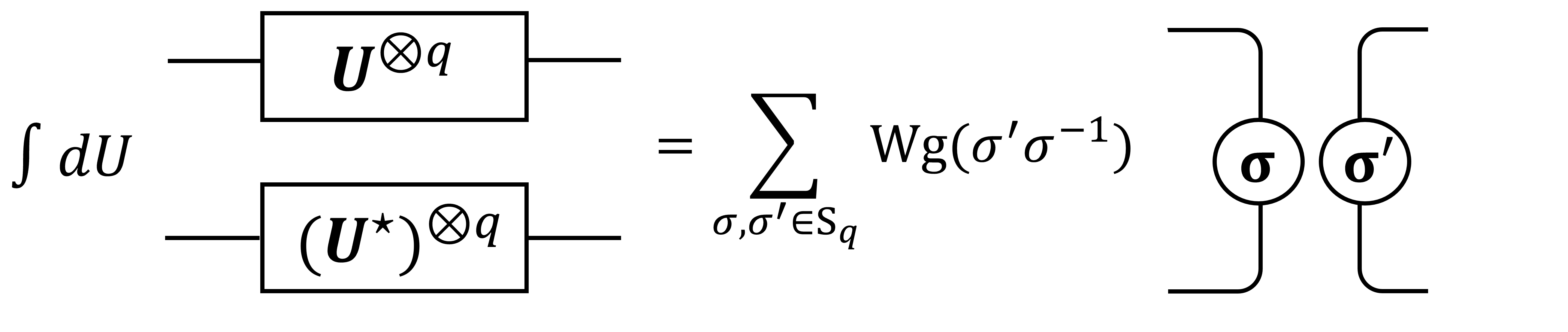}
\label{eq:weingarten}
\end{equation}
The ${\rm Wg}(\sigma \sigma'^{-1})$ are rational functions of the dimension of the underlying Hilbert space $EN$. In the diagrammatic expression above, we have taken the transpose of $U^\dagger$ in order to align indices in the same type of Hilbert space---$EN$ on one side,  $RB'$ on the other side. 

The permutations $\sigma$ and $\sigma'$ switch around the copies of $EN$ and $RB'$. For example, for $q=3$, the permutation $(12)(3) \in S_3$ maps the basis element 
\begin{equation}
| i \rangle_{(EN)_1} \otimes | j \rangle_{(EN)_2} \otimes | k \rangle_{(EN)_3}
~\xrightarrow{~(12)(3)~}
| j \rangle_{(EN)_1} \otimes | i \rangle_{(EN)_2} \otimes | k \rangle_{(EN)_3}
\end{equation}
Diagrammatically, the legs of the $q$ copies of $U$ are identified with the corresponding legs of the $q$ copies of $U^*$, with a pattern set by $\sigma \in S_q$ (for column indices) and by $\sigma' \in S_q$ (for row indices).

Substituting (\ref{eq:weingarten}) into (\ref{eq:oldstructure}) reduces the calculation to a product of loops. Each loop contributes one factor of the size of its Hilbert space---that is $N$, $E$, $R$, or $B'$. Therefore, the computation requires us to enumerate loops of each type, as a function of $\sigma, \sigma' \in S_q$. 

Each loop in question is formed by joining lines that arise from (\ref{eq:weingarten}) with legs, which are shown in (\ref{eq:oldstructure}). The latter legs likewise connect to one another copies of the same vector space on $U$ and $U^*$. Therefore, they too effectively define permutations on $S^q$ as follows:
\begin{align}
\pi_{B'} = \pi_{N} & = \iota = \textrm{identity} \\
\pi_E = \pi_R & = \xi = \textrm{generator~of~}\mathbb{Z}_q
\label{pierb}
\end{align}
It is easy to understand these permutations in terms of the structure of ${\rm tr}\,(\rho_{ER})^q$. $B'$- and $N$-legs are contracted within-copy to form $\rho_{ER}$ whereas $E$ and $R$ are contracted cyclically across copies, so as to form the product of $q$ copies of $\rho_{ER}$'s. 

We are therefore needing to count loops, which arise when the permutations from the average over unitaries ($\sigma, \sigma'$) get contracted with $\iota$ and $\xi$---the permutations defined by the structure of ${\rm tr}\,(\rho_{ER})^q$. Given two general permutations $\alpha$ and $\beta$, the number of cycles in $\alpha \beta^{-1}$ is called $l(\alpha, \beta)$; this is the quantity we need. Putting this all together, we find:
\begin{equation}
{\rm tr}\,(\rho_{ER})^q = 
(EN)^{-q} \sum_{\sigma, \sigma' \in S_q} {\rm Wg}(\sigma \sigma'^{-1}) 
B'^{l(\sigma, \iota)} R^{l(\sigma, \xi)} N^{l(\sigma', \iota)} E^{l(\sigma', \xi)}
\label{trrhoqfull}
\end{equation}
The factor $(EN)^{-q}$ normalizes $q$ copies of maximally entangled states on $EB$ and on the diary-reference pair $NA$.

\paragraph{Exact example: $q=2$} The summations over $\sigma, \sigma' \in S_2$ give four terms. Two of them, coming from $\sigma \sigma'^{-1} = \iota$, are weighted by ${\rm Wg}(\iota) = 1/\big((EN)^2-1\big)$. For the two others we have $\sigma \sigma^{-1} = \xi$ and the Weingarten factor is ${\rm Wg}(\xi) = 1/EN\big((EN)^{-2}-1\big)$. Collecting all terms, we obtain:
\begin{align}
{\rm tr}\,(\rho_{ER})^2 = 
&\, (EN)^{-2}\,
\frac{B'^{l(\iota, \iota)} R^{l(\iota, \xi)} N^{l(\iota, \iota)} E^{l(\iota, \xi)} 
+ B'^{l(\xi, \iota)} R^{l(\xi, \xi)} N^{l(\xi, \iota)} E^{l(\xi, \xi)}}{(EN)^2-1} 
\nonumber \\
+ &\, (EN)^{-2} \,
\frac{B'^{l(\iota, \iota)} R^{l(\iota, \xi)} N^{l(\xi, \iota)} E^{l(\xi, \xi)} 
+ B'^{l(\xi, \iota)} R^{l(\xi, \xi)} N^{l(\iota, \iota)} E^{l(\iota, \xi)}}{EN \big((EN)^2-1\big)}
\label{trrho2full}
\end{align}
Substituting the numbers of cycles $l(\iota,\iota) = l(\xi,\xi) = 2$ and $l(\iota, \xi) = 1$, we get:
\begin{align}
{\rm tr}\,(\rho_{ER})^2 
& = \frac{B' N^2 E + R N E^2 + B' E + R N}{EN\big((EN)^2-1\big)} \\
\Longrightarrow \quad S^{(2)}(\rho_{ER})
& = -\log \frac{B' N^2 E + R N E^2 + B' E + R N}{EN\big((EN)^2-1\big)}
\end{align}
After recalling that $S^{(q)}(\rho_{ERN}) = b' = e-c$ and $S^{(2)}(\rho_{N}) = n$, we find:
\begin{align}
I^{(2)} (ER:N) 
& = -\log \frac{B' N^2 E + R N E^2 + B' E + R N}{EN\big((EN)^2-1\big)} + n+c - e
\nonumber \\
& = 2n-\log \frac{1 + \exp(-2c) + \exp(-2n-2c) + \exp(-2e)}{1 + \exp(-2e-2n)} 
\end{align}
This is the exact average over $U \in U(EN)$. In the limit of interest---where $E, B' \gg R, N$---the dominant contribution comes from the term $\sigma = \sigma' = \xi$ in equation~(\ref{trrho2full}). The leading correction comes from $\sigma = \sigma' = \iota$. Assuming the overhead $c \gg 1$, we have:
\begin{equation}
I^{(2)} (ER:N) =  2n-\exp(-2c)+\ldots 
\end{equation}

\paragraph{Black hole limit $E, B' \gg R, N$ for integer $q$} Equation~(\ref{trrhoqfull}) in principle calculates the exact $U(EN)$-average of $S^{(q)}(\rho_{ER})$ and $I^{(q)} (ER:N)$ for all integer $q$. We would like to simplify this expression in the limit of physical interest where $E, B' \gg R, N$.

Each summand in (\ref{trrhoqfull}) is labeled by a pair of permutations in $S_q$. The task is to characterize those $\sigma$ and $\sigma'$, which have the fastest scaling with $e$ and $b' = e-c$. To identify the dominant terms, we inspect the factor $B'^{l(\iota, \sigma)} {\rm Wg}(\sigma \sigma'^{-1}) E^{l(\sigma', \xi)}$. 

To enable a direct comparison between terms, it is useful to understand the Weingarten factor in terms of the cycle-counting function. In the limit of large Hilbert space dimension, its leading order behavior is:
\begin{equation}
{\rm Wg}(\sigma \sigma'^{-1}) = 
(EN)^{-2q + l(\sigma, \sigma')} {\rm wg}(\sigma \sigma'^{-1})
\Big(1 + \mathcal{O}\big((EN)^{-2}\big) \Big)
\label{wgscaling}
\end{equation}
The combinatorial coefficient ${\rm wg}(\sigma \sigma'^{-1})$ has no dependence on Hilbert space dimension, only on the cycle structure of $\sigma \sigma'^{-1} \in S_q$. We conclude that the leading terms in (\ref{trrhoqfull}) are those which maximize the combination:
\begin{equation}
l(\iota, \sigma) + l(\sigma, \sigma') + l(\sigma', \xi)
\label{tomax-large}
\end{equation}
It is easier to understand this expression in terms of the Cayley distance
\begin{equation}
d(\sigma, \sigma') = q - l(\sigma, \sigma')
\end{equation}
Because a $p$-cycle decomposes into $p-1$ transpositions, $d(\sigma, \sigma')$ gives the minimal number of transpositions required to generate $\sigma \sigma'^{-1}$. An important property of $d(.\,, .)$ is the triangle inequality: $d(\alpha, \beta) + d(\beta, \gamma) \geq d(\alpha, \gamma)$.

In terms of $d(.\,,.)$, to maximize (\ref{tomax-large}) is to minimize:
\begin{equation}
d(\iota, \sigma) + d(\sigma, \sigma') + d(\sigma', \xi)
\label{tomin-large}
\end{equation}
By the triangle inequality, this is at least $d(\iota, \xi) = q-1$; we are looking to saturate this lower bound. In geometric parlance, we need $\sigma$ and $\sigma'$ which are `colinear' with $\iota$ and $\xi$ in the metric $d(.\,,.)$, that is they live on a shortest path from $\iota$ to $\xi$ with respect to the Cayley distance.

Such $\sigma$ and $\sigma'$ are easily identified using facts, which were reviewed in \cite{akers2022reflected}. The requirement that $\sigma$ and $\sigma'$ live on a shortest path from $\iota$ to $\xi$ means that they are non-crossing permutations: $\sigma, \sigma' \in {\rm NC}_q$. After that, we must still ensure that they live on \emph{the same} shortest path. Permutations $\sigma$, which can be passed on a shortest path from $\iota$ to a given permutation $\sigma'$, were also characterized in \cite{akers2022reflected}. We repeat that characterization below, subject to two changes---one in the notation, and one in restricting to $\sigma' \in {\rm NC}_q$:
\medskip

\noindent
\underline{Corollary~5 in Appendix~A of \cite{akers2022reflected}} (specialized to $\sigma' \in {\rm NC}_q$):  \emph{Suppose $\sigma' \in {\rm NC}_q$ decomposes into disjoint cycles $C_1, C_2, . . . , C_k$. Then $\sigma$ lies on a geodesic between $\iota$ and $\sigma'$ if and only if $\sigma \in {\rm NC}_q$ and each $C_j$ is a union of disjoint cycles of $\sigma$.}
\medskip

\noindent
We will denote this condition $\sigma \leq \sigma'$. We have $\iota \leq \sigma$ and $\sigma' \leq \xi$ as obvious special cases of this notation.\footnote{We could have used the unrestricted version of the Corollary to define the relation $\sigma \leq \sigma'$, without stipulating in the definition of $\leq$ that $\sigma' \in {\rm NC}_q$. Doing so would allow us to write the sum in equation~(\ref{trrholargee}) simply as $\sum_{\iota \leq \sigma \leq \sigma' \leq \xi}$, which follows directly from equation~(\ref{tomin-large}). In that approach, $\sigma \in {\rm NC}_q$ (and likewise for $\sigma'$) \emph{because} this is equivalent to $\iota \leq \sigma \leq \xi$. That way of proceeding is more elegant but it may be less transparent in a quick reading.} 

We now truncate all terms, which are subleading in $E$. Organizing all the remaining terms in powers of $N = \exp(n)$ and $R/N = \exp(c)$, we obtain: 
\begin{align}
{\rm tr}\,(\rho_{ER})^q & = 
(EN^2/R)^{1-q}
\!\!\!\sum_{\substack{\sigma, \sigma' \in {\rm NC}_q\\ \sigma \leq \sigma'}} \!\!\!
{\rm wg}(\sigma \sigma'^{-1})
N^{-2d(\sigma', \sigma)}
(R/N)^{-2d(\sigma, \xi)} 
\label{trrholargee} \\
S^{(q)}(\rho_{ER}) & = e + n - c - \frac{1}{q-1} 
\,\log\!\!\! \sum_{\substack{\sigma, \sigma' \in {\rm NC}_q\\ \sigma \leq \sigma'}} \!\!\!\!
{\rm wg}(\sigma \sigma'^{-1})
\exp\!\Big(\!-\!2\big[n \times d(\sigma', \sigma) + c \times d(\sigma, \xi)\big]\Big)
\label{iqlargee}
\end{align}
We have used the colinearity $\iota \leq \sigma \leq \sigma' \leq \xi$ in simplifying the exponents of $N$ and $R/N$.

If the remaining parameters of the problem---$n$ and $c$---do not form a hierarchy, expression~(\ref{iqlargee}) does not simplify further. To make progress, we assume $n \gg c$. That is, in attempting to read Alice's diary, we will use an overhead that is parametrically smaller than the diary itself. It would make no sense to consider $c$ parametrically smaller than 1, so our assumption also implies $n \gg 1$. 

This regime is dominated by terms with $\sigma = \sigma'$, with a single extant sum over $\sigma \in {\rm NC}_q$:
\begin{equation}
S^{(q)}(\rho_{ER}) = e + n - c - \frac{1}{q-1} 
\,\log\!\sum_{\sigma \in {\rm NC}_q}
\exp\!\big(\!-\!2 c \times d(\sigma, \xi)\big)
\label{iqlargeen}
\end{equation}
Going from (\ref{iqlargee}) to (\ref{iqlargeen}) uses ${\rm wg}(\iota) = 1$. The sum can be evaluated exactly by using $d(\sigma, \xi) = l(\sigma, \iota) - 1$ and the fact that the number of non-crossing partitions with exactly $l$ cycles is the Narayana number
\begin{equation}
N(q,l) = \frac{1}{q}\binom{q}{l}\binom{q}{l-1}.
\end{equation}
In the end, the sum in~(\ref{iqlargeen}) evaluates to the hypergeometric function and gives:
\begin{equation}
S^{(q)}(\rho_{ER}) = e + n - c - \frac{1}{q-1} \log \phantom{.}_2F_1(1-q, -q; 2, e^{-2c})
\label{iqlargeenfinal}
\end{equation}
These quantities---which we believe have not been calculated before---are relevant when the overhead of captured radiation over diary size, $c = r-n$, is not parametrically large.

If it is large, we Taylor-expand the hypergeometric function:
\begin{equation}
\phantom{.}_2F_1(1-q, -q; 2, e^{-2c}) = 1 + e^{-2c} q(q-1)/2 + \mathcal{O}(e^{-4c})
\end{equation}
In reference to equation~(\ref{iqlargeen}), the leading term comes from $\sigma = \xi$ whereas the first subleading term comes from non-crossing permutations $\sigma$ that have precisely two cycles, of which there are $q(q-1)/2$. Substituting the Taylor expansion yields the $e \gg n \gg c \gg 1$ answer:
\begin{equation}
S^{(q)}(\rho_{ER}) \approx e + n - c - \frac{q}{2} \exp(-2c)
\label{iqlargeenc}
\end{equation}
After plugging in $S^{(q)}(\rho_N) = n$ and $S^{(q)}(\rho_{ERN}) = S^{(q)}(\rho_{B'}) = b' = e + n - r = e-c$, we obtain the R{\'e}nyi mutual informations:
\begin{align}
I^{(q)}(ER:N) & = 2n - (q/2)\, e^{-2c} \label{iqernold} \\
I^{(q)}(ER:B') & = 2b' - (q/2)\, e^{-2c} \\
I^{(q)}(N:B') & = \phantom{2b'} + (q/2)\, e^{-2c} \label{iqnbold}
\end{align}
Ordinary (non-R{\'e}nyi) mutual informations are given by their $q \to 1$ limit. Naturally, these quantities confirm that both $N$ and $B'$ are maximally entangled with the aggregate radiation $ER$, modulo exponentially small corrections.

\subsection{Young black hole}
\label{sec:young}
We now turn to black holes, which are not yet fully entangled with the early radiation. The microstates of such black holes form a Hilbert space $B$. In the previous subsection $B$ did not make an independent appearance; we treated $E$ as a proxy for $B$. We now assume they are distinct with $B > E$.

The initial black hole-radiation system is prepared in some initial state $| \psi \rangle \in BE$, which we again treat as random. We continue to assume $b \gg n + c = r$ but do not assume a hierarchy between $e$ and $n, c$.

The introduction of $| \psi\rangle$ does not change the calculation of $\rho_N = 1/N$. It does, however, change the structure of $\rho_{ER}$ and $\rho_{ERN}$. We begin with $\rho_{ER}$ because it forms a sharp contrast with the previous subsection. 

\paragraph{R{\'e}nyi entropies of $\rho_{ER}$} In comparison to equation~(\ref{eq:oldstructure}), we need to insert the outer product $| \psi\rangle \langle \psi |$ on every consecutive pair of $E$-lines. This modification produces the following structure for ${\rm tr}\,(\rho_{ER})^q$:
\begin{equation}
\centering
\includegraphics[width=0.8\textwidth]{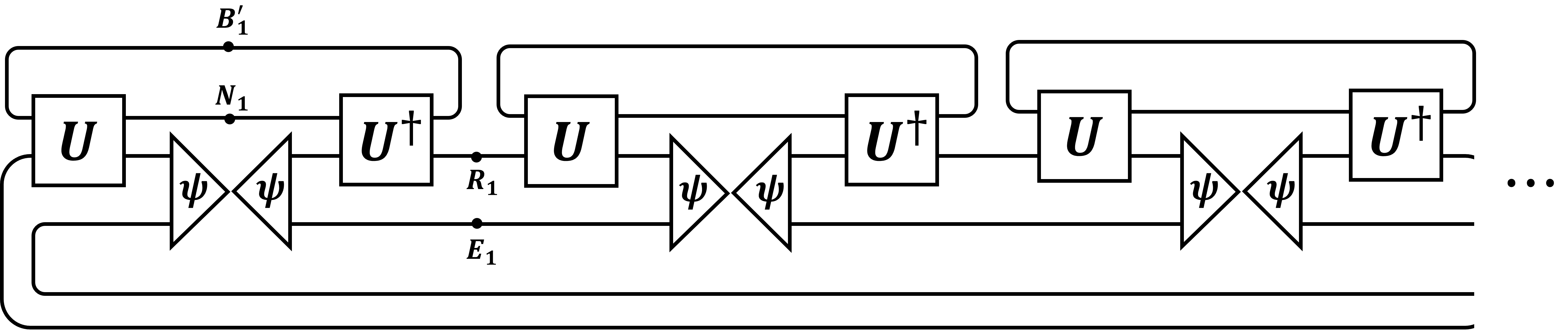}
\label{eq:youngstructure}
\end{equation}
To compute this, we need the averaging relation (see also Appendix~\ref{sec:weingarten}):
\begin{align}
\overline{|\psi\rangle\langle\psi|^{\otimes q}} 
& = \left[(BE)^{q}f(BE)\right]^{-1} \sum_{\tau\in S_{q}} \Pi_{\tau} 
\label{eq:averagepsiq} \\ 
f(BE) 
& = (BE)^{1-q} \frac{(BE+q-1)!}{(BE)!}
=(BE)^{-q} \sum_{\tau\in S_{q}}(BE)^{l(\tau,\iota)} 
=1 + \frac{q(q-1)}{2BE} + \ldots
\label{deffnorm}
\end{align}
The $\Pi_\tau$ are permutation matrices, which permute the $q$ copies of $|\psi\rangle$ relative to $\langle \psi|$. Both middle expressions for the normalization are exact, but we will only use the large $BE$ expansion. In graphical notation, equation~(\ref{eq:averagepsiq}) reads:
\begin{equation}
\centering
\includegraphics[width=0.8\textwidth]{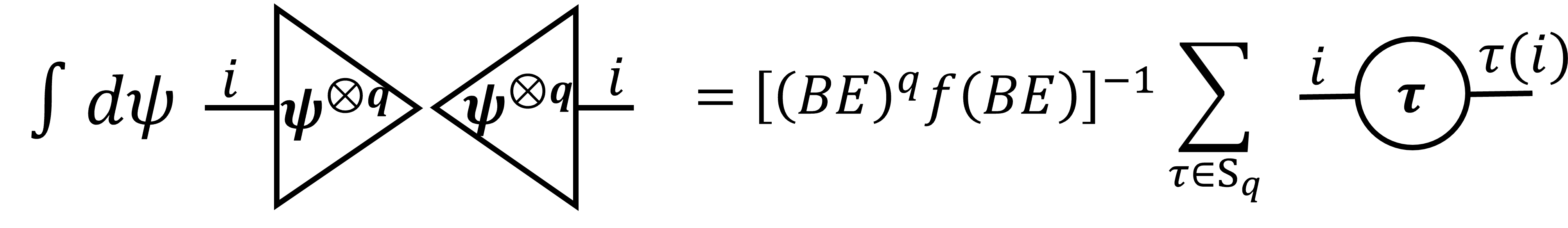}
\end{equation}

The insertion of $| \psi\rangle \langle \psi |$ has a simple effect on ${\rm tr}\,(\rho_{ER})^q$. $E$-loops of equation~(\ref{trrhoqfull}) now split into $E$-loops and $B$-loops. $B$-loops are formed by contracting $\tau \in S_q$ with $\sigma'$ from the $UU^\dagger$ average; there are $l(\sigma', \tau)$ of them. $E$-loops are formed by contraction with $\pi_B = \xi$ (see equation~\ref{pierb}); there are $l(\tau, \xi)$ of them. In the end, we find
\begin{equation}
{\rm tr}\,(\rho_{ER})^q = 
[N^q (BE)^qf(BE)]^{-1} \!\!\sum_{\tau,\sigma, \sigma'\in S_q} {\rm Wg}(\sigma \sigma'^{-1}) 
B'^{l(\iota, \sigma)} R^{l(\sigma, \xi)} N^{l(\iota, \sigma')} B^{l(\sigma', \tau)} E^{l(\tau, \xi)}
\label{trrhoqfullyoung}
\end{equation}
for the exact $U(BN)$ and $|\psi\rangle$-average. Note that the factor $[(BE)^qf(BE)]^{-1}$ replaces the normalization factor $E^{-q}$ in equation~(\ref{trrhoqfull}).

We now assume that $B$ dominates over all other independent parameters. In particular, this assumption entails a hierarchy $B \gg E$. The `technically young but actually middle-aged' regime where $B \gtrsim E$ is briefly discussed in the next subsection.

Let us identify terms leading in $B$ and $B' = BN/R$. Using~(\ref{wgscaling}) and the fact that $f(BE)$ is of order unity, we see that the leading exponent of $B$ comes from maximizing
\begin{equation}
-3q + l(\iota, \sigma) + l(\sigma, \sigma') + l(\sigma', \tau) 
= - d(\iota, \sigma) - d(\sigma, \sigma') - d(\sigma', \tau)
\label{btomax}
\end{equation}
In contrast to (\ref{tomin-large}), it is maximized by $\sigma = \sigma' = \tau = \iota$ at zero. The leading corrections are of order $1/B$. A priori, they can come either from (i) the first subleading term in $1/f(BE)$ or from cases where (ii) $\tau$ alone or (iii) $\sigma' = \tau$ or (iv) $\sigma = \sigma' = \tau$ is a single transposition.\footnote{There are also corrections from the parenthesis in equation~(\ref{wgscaling}), but they are of relative order $E^{-2}$ or $B^{-2}$. As such, they never enter the leading or subleading order in any calculation in this paper.}  In all four cases, these terms come with a factor of $q(q-1)/2$, which counts transpositions. Writing all $\mathcal{O}(B^{-1})$ corrections explicitly, we find:
\begin{equation}
{\rm tr}\,(\rho_{ER})^q = (ER)^{1-q} \left( 1 + \frac{q(q-1)}{2B} 
\left(-\frac{1}{E} + E - \frac{E}{N^2} + \frac{ER^2}{N^2}\right) + \mathcal{O}\left( \frac{1}{B^2} \right) \right)
\label{rhoerq}
\end{equation}
Correction terms are listed in order (i-iv). In term (iii) we used $\rm{wg(transposition)} = -1$. This calculation implies $\rho_{ER} \approx 1/ER$.

Because $R/N = \exp(c) > 1$, the largest of the four correction terms is $ER^2 / BN^2$. Assuming $b-e \gg c \gg 1$ we find
\begin{equation}
S^{(q)}(\rho_{ER}) = e + r - \frac{q}{2} \exp( -b + e + 2c)
\label{erq}
\end{equation}
This result also identifies the way to reverse the finding that $ER$ is nearly maximally mixed: to set $b \sim e + 2c$ or $R \sim N\sqrt{B/E}$. In other words, the overhead radiation to be collected (over and above diary size) must be half as large as the entropy gap $(b-e)$, which separates the black hole from being old at the time Alice tosses her diary. 

\begin{figure}[t]
\centering
\includegraphics[width=0.98\textwidth]{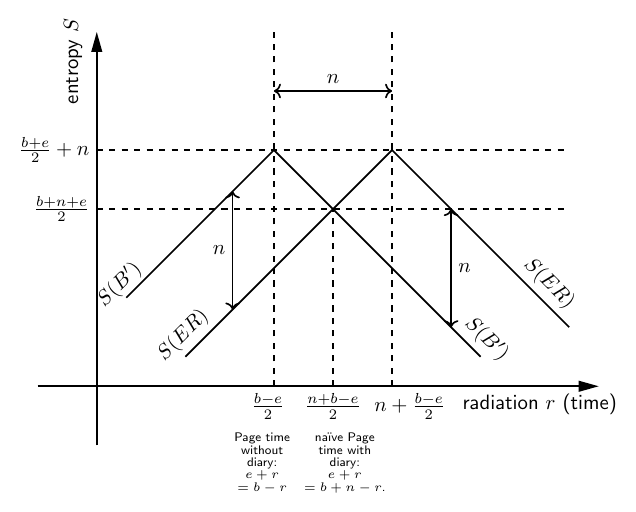}\\
\caption{The effect of dropping a diary into a young black hole on entanglement entropies $S(\rho_{ERN}) = S(\rho_{B'})$ and $S(\rho_{ER})$, derived from equations~(\ref{erq}) and (\ref{ernq}). We mark the range of radiation (\ref{rrange}) when the pairwise mutual informations of $ER$, $N$ and $B'$ are all non-zero. $ER$ ceases to be maximally mixed later than the putative Page time, so dropping the diary when the black hole is still young delays the mirroring of information.}
\label{fig:entropies}
\end{figure}

Let us first inspect this conclusion at $n = 0$, that is when Alice does not toss a diary at all. Say we collect $c/\log 2 = (b-e)/2 \log 2$ extra bits of radiation. If we adjoin them to the early radiation $E$, we effectively reset $e \to e_{\rm shifted} = e+(b-e)/2 = (b+e)/2$. Meanwhile, the black hole shrinks to size $b_{\rm shift} = b - (b-e)/2 = (b+e)/2$ and becomes old, which is the expected conclusion. Now reintroduce the diary: $n > 0$. The radiation that must be collected to avert $\rho_{ER} \approx (ER)^{-1}$ jumps to $n + (b-e)/2$. In other words, Alice's tossing of the diary delays the threshold when the radiation ceases to be maximally mixed. It does so by an amount, which---in entropic terms---equals the size of the diary. These conclusions are illustrated in Figure~\ref{fig:entropies}.

\paragraph{R{\'e}nyi entropies of $\rho_{ERN}$}  Quantity ${\rm tr}\,(\rho_{ERN})^q$ has the following diagrammatic representation:
\begin{equation}
\centering
\includegraphics[width=0.8\textwidth]{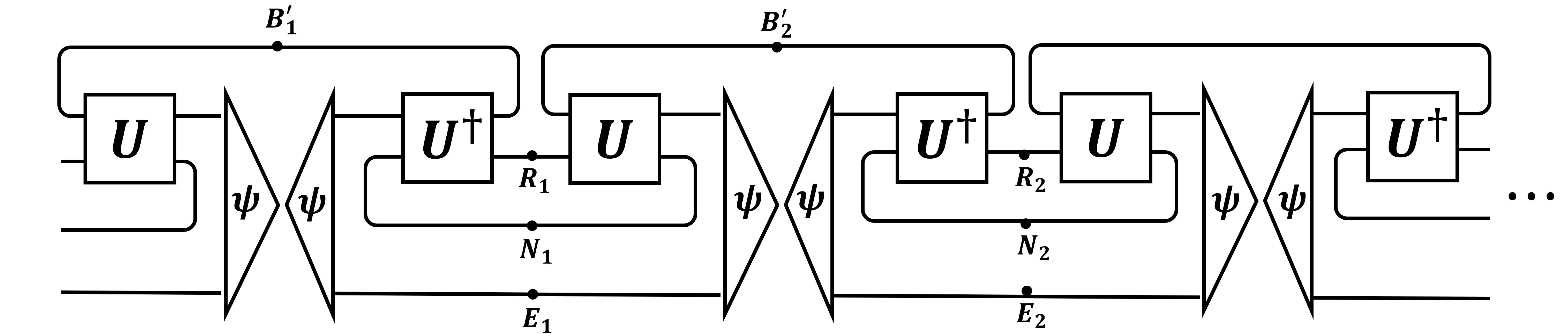}
\end{equation}
In comparison with equation~(\ref{trrhoqfullyoung}), $N$-loops are formed by contraction with $\xi$ instead of $\iota$ because $N$-legs are contracted across copies, not within-copy:
\begin{equation}
{\rm tr}\,(\rho_{ERN})^q = 
[N^q (BE)^q f(BE)]^{-1} \!\!\sum_{\tau,\sigma, \sigma'\in S_q} {\rm Wg}(\sigma \sigma'^{-1}) 
B'^{l(\iota, \sigma)} R^{l(\sigma, \xi)} N^{l(\xi, \sigma')} B^{l(\sigma', \tau)} E^{l(\tau, \xi)}
\label{trrhonfullyoung}
\end{equation}
This change does not affect the counting of powers of $B$, which we assume dominate the calculation. The logic spelled out in and below equation~(\ref{btomax}) carries over and we find:
\begin{equation}
{\rm tr}\,(\rho_{ERN})^q = (ERN)^{1-q} \left( 1 + \frac{q(q-1)}{2B} 
\left(-E^{-1} + E - E + ER^2\right) + \mathcal{O}\left( B^{-2} \right) \right)
\label{eq:rho_ERN_young}
\end{equation}
We have organized the correction terms the same way as in equation~(\ref{rhoerq}). After dropping terms other than $ER^2/B$, the R{\'e}nyi entropies become:
\begin{equation}
S^{(q)}(\rho_{ERN}) = e + 2n + c - \frac{q}{2} e^{-(b-e) + 2(n+c)}
\label{ernq}
\end{equation}
The conclusion that $\rho_{ERN} \approx (ERN)^{-1}$ becomes unreliable when $R \sim \sqrt{B/E}$. What happens at that point? To answer this, note that $ERN$ is the purifier of $B'$ whose maximal entanglement is set by $B' = BN / R$. The correction term $ER^2/B$ is a harbinger of the old black hole bound $S(\rho_{B'}) \leq b' = b+n-r$. It becomes order one when this bound coincides with $S(\rho_{ERN}) \leq e + r +n$, i.e. when $ERN = BN/R$.

\paragraph{Mutual information} Combining the calculations, we find the following:
\begin{align}
I^{(q)}(ER:N) & = \phantom{2(e+r)} + (q/2)\, e^{ -(b - e) + 2r}
\label{iqernyoung} \\ 
I^{(q)}(ER:B') & = 2(e+r) - (q/2)\, e^{ -(b - e) + 2r} \\
I^{(q)}(N:B') & = 2n \phantom{2(+r.} - (q/2)\, e^{ -(b - e) + 2r} 
\label{iqnbyoung}
\end{align}
For standard mutual informations set $q \to 1$. These expressions are valid for $b \gg r \gg n \gg 1$ and $b \gg e$. They say that both $ER$ and $N$ are, up to exponentially small corrections, maximally entangled with the remaining black hole $B'$. 

Each of the three mutual informations---$I(ER:N)$, $I(ER:B')$ and $I(N:B')$---transitions between distinct values as the black hole radiates and becomes old. Following our discussion below equations~(\ref{erq}) and (\ref{ernq}), the transition happens over a range of radiation sizes; see Figure~\ref{fig:entropies}.
\begin{equation}
\sqrt{B/E} \lesssim R \lesssim N\sqrt{B/E}
\label{rrange}
\end{equation}
The lower range is when we switch the active constraint from $S(\rho_{B'}) \leq e + n + r$ to $S(\rho_{B'}) \leq b + n - r$. The upper range is when $S(\rho_{ER})$ transitions from $e + r = e + n + c$ to $b + n - c$. Both transitions are clearly heralded by $\mathcal{O}(B^{-1})$ correction terms becoming order one. 

We close with some intriguing observations, which are most neatly expressed using entropic (lowercase) parameters: 
\begin{itemize}
\item The range of radiation over which the transition occurs is set by the size of Alice's diary: $\Delta r \approx n$.
\item The midpoint of the transitional range occurs when the black hole, with Alice's diary inside it, becomes old: $b + n - r_{\rm mid} = e + r_{\rm mid}$, equivalent to $R_{\rm mid} \sim \sqrt{BN/E}$. 
\item By tossing her diary into the black hole, Alice has delayed by $n = \log N$ the Page time, as defined by the approximately maximally mixed character of $\rho_{ER}$. 
\end{itemize}

The next subsection explores the regime when the `Page gap' $b-e$ is not large. This allows us to relate and contrast the calculations in Section~\ref{sec:old} and \ref{sec:young}.

\subsection{Technically young but middle aged black hole} 
\label{sec:midage}
It is interesting to inspect a neighborhood of the Page time and see how the R{\'e}nyi entropies interpolate between the answers in Sections~\ref{sec:old} and \ref{sec:young}. With our technology, we can do so by studying the regime $B > E$ but not $B \gg E$. 

\paragraph{R{\'e}nyi entropies of $\rho_{ER}$} Since the black hole is technically young, we start with expression (\ref{trrhoqfullyoung}). However, we now look for terms leading in $B$, $B'$ and $E$. Rather than (\ref{btomax}), we are needing to maximize 
\begin{equation}
-d (\iota, \sigma) - d(\sigma, \sigma') - d(\sigma', \tau) - d(\tau, \xi)
\label{midagemax}
\end{equation}
Because now $B \not\gg E$, the factor $E^{l(\tau, \xi)}$ reintroduces a penalty for $\sigma, \sigma', \tau$ being far from $\xi$ in Cayley distance. In the end, we find that $\sigma, \sigma', \tau$ must lie on a `geodesic' from $\iota$ to $\xi$ in metric $d(.,\,.)$. This makes the calculation nearly identical to (\ref{trrholargee}) in the old black hole regime. The only differences are: 
\begin{itemize}
\item Since now $E$ and $B$ are distinct and it is $B$ that the random unitary $U$ acts on, we replace $E \to B$.
\item We have a triple sum over $\sigma \leq \sigma' \leq \tau$, which are sandwiched between $\iota \leq \ldots \leq \xi$.
\item Normalization: ${\rm tr}\,(\rho_{ER})^q$ gets an extra factor of $[E^q f(BE)]^{-1}$.
\item We replace $B^{l(\sigma', \xi)} \to B^{l(\sigma', \tau)} E^{l(\tau, \xi)} 
= E^q B^{l(\sigma', \xi)} (B/E)^{d(\tau, \xi)}$; the equality follows from colinearity $\sigma' \leq \tau \leq \xi$. Then factor $E^q$ cancels against one in the normalization.
\end{itemize}
Applying these changes to equation~(\ref{trrholargee}), we obtain:
\begin{equation}
{\rm tr}\,(\rho_{ER})^q = 
\frac{(B'N)^{1-q}}{f(BE)}
\!\!\!\sum_{\substack{\sigma,\, \sigma', \, \tau \in {\rm NC}_q\\ \sigma \leq \sigma' \leq \tau}} \!\!\!
{\rm wg}(\sigma \sigma'^{-1})
(B/E)^{d(\tau, \xi)}
N^{-2d(\sigma', \sigma)}
(R/N)^{-2d(\sigma, \xi)} 
\label{trrholargee-midage} 
\end{equation}
In the overall factor we used $B' = BN/R$. 

If we do not have a hierarchy $B / E \gg 1$, equation~(\ref{trrholargee-midage}) behaves exactly like (\ref{trrholargee}) in Section~\ref{sec:old}; it only differs from it by factors of order unity. This is consistent with our conclusion below equation~(\ref{erq}): $B/E$ sets a scale for the size of the diary, past which $ER$ moves away from being nearly maximally mixed. Here we are looking at the regime where $B/E \sim 1$ so $ER$ ceases to be nearly maximally mixed right away.

\paragraph{R{\'e}nyi entropies of $\rho_{ERN}$} 
In contrast to Section~\ref{sec:old}, $\rho_{B'}$ is not maximally mixed. How does $\rho_{B'}$ transition between the two regimes? 

We again start with the young black hole answer in equation~(\ref{trrhonfullyoung}). Looking first at powers of $B$ and $E$, the maximal collective exponent of $B$ is obtained if $\sigma' \leq \tau \leq \xi$ are colinear, in which case $B^{l(\sigma', \tau)} E^{l(\tau, \xi)} = E^q B^{l(\sigma', \xi)} (B/E)^{d(\tau, \xi)}$. Substituting this in (\ref{trrhonfullyoung}), we obtain:
\begin{equation}
{\rm tr}\,(\rho_{ERN})^q =
[(BN)^q f(BE)]^{-1} \!\!\!\!\sum_{\sigma, \sigma'\in S_q}
\!\!\! {\rm Wg}(\sigma \sigma'^{-1}) B'^{l(\iota, \sigma)} R^{l(\sigma, \xi)} (BN)^{l(\sigma', \xi)}  
\!\!\sum_{\tau \geq \sigma'} (B/E)^{d(\tau, \xi)}
\label{trrhonfullmidage}
\end{equation}
Now the exponent of $B$ and $B'$ to be maximized is:
\begin{equation}
-q + \big(-2q + l(\sigma, \sigma')\big) + l(\iota, \sigma) + l(\sigma', \xi)
= -d(\iota, \sigma) - d(\sigma, \sigma') - d(\sigma', \xi)
\label{maxmidage}
\end{equation}
Once again, setting $B \not\gg E$ has reintroduced a penalty for $\sigma, \sigma'$ being far from $\xi$ in Cayley distance. It forces the leading terms to lie on a geodesic from $\iota$ to $\xi$, where the exponent of $B$ and $B'$ is $1-q$. After truncating terms subleading in $B$, substituting $B' = BN/R$ and simplifying, we find:
\begin{equation}
{\rm tr}\,(\rho_{ERN})^q = 
\frac{(B')^{1-q}}{f(BE)}
\sum_{\sigma \in {\rm NC}_q} 
R^{-2d(\sigma, \xi)} 
\left(
\sum_{\sigma' \geq \sigma} {\rm wg}(\sigma \sigma'^{-1})
\sum_{\tau \geq \sigma'} (B/E)^{d(\tau, \xi)} \right)
\label{trrhonfullmidage2}
\end{equation}
If $R \gg 1$, the leading term comes from $\sigma = \xi$, in which case maximizing (\ref{maxmidage}) also sets $\sigma'=\tau=\xi$ and the factor in parenthesis is unity. The first subleading term comes from $\sigma$'s, which are one transposition away from $\xi$; there are $q(q-1)/2$ of them. But for those $\sigma$'s, the sum in parentheses is some number, which is order unity but not equal to 1; call it $\chi(B/E)$. Therefore, retaining only first subleading corrections, we get: 
\begin{align}
{\rm tr}\,(\rho_{ERN})^q & = (B')^{1-q} \left( 1 + \frac{q(q-1)}{2}\frac{\chi(B/E)}{R^2} + \ldots\right) 
\label{tracemidage} \\
S^{(q)}(\rho_{ERN}) & = b' - \frac{q}{2}\, e^{-2r} \chi(B/E) 
\end{align}

How to understand the correction term? Notice that setting $\tau = \xi$ in (\ref{trrhonfullmidage}) returns to the settings of the old black hole, where all $U$ and $U^\dagger$s cancel out and $\rho_{B'} = 1/B'$. Thus, the corrections come entirely from $\tau \neq \xi$. Even when we go to Page time at $B=E$, the $\tau \neq \xi$ continue to contribute nonvanishingly. Therefore, at Page time, we can understand these corrections as quantifying the difference between the maximally mixed initial state on $EB$, which we assumed in Section~\ref{sec:old}, and a Haar-random state on $EB$ assumed here. In the end, the correction term in (\ref{tracemidage}) is that dictated by Page's theorem \cite{Page1993average}.

\section{Reflected R{\'e}nyi entropies}
\label{sec:reflected}

Our computations can be adapted to more general quantities, other than R{\'e}nyi entropies. In this section we compute reflected R{\'e}nyi entropies, which generalize the reflected entropy defined by Dutta and Faulkner~\cite{dutta2021canonical}. For a density matrix $\rho_{XY}$ describing a mixed state on Hilbert space $\mathcal{H}_{X}\otimes\mathcal{H}_Y$, the reflected entropy $S_R(X:Y)$ quantifies correlations between $X$ and $Y$. 

This quantity has been extensively studied in the realm of quantum information theory~\cite{akers2020entanglement,hayden2021markov,zou2021universal,Hayden2023notAcorrelationMeasure},
many-body systems~\cite{kusuki2020entanglement,kudler2020correlation,moosa2020time,bueno2020reflected-1,berthiere2021topological,kudler2021quasi,bueno2020reflected-2,camargo2021long, dutta2022reflected}, 
and holographic models~\cite{dutta2021canonical,bao2019multipartite,chu2020generalizations,bao2021entanglement,akers2022reflected,akers2022reflected-2,basu2022entanglement,akers2020entanglement}.
Especially in holographic duality the reflected entropy holds much interest because it is conjectured to be dual to the cross-section of an entanglement wedge \cite{dutta2021canonical}. Supporting evidence from random tensor network (RTN) states \cite{Hayden2016Holographic} includes \cite{akers2022reflected, akers2022reflected-2}. Because product states $\rho_X \otimes \rho_Y$ have entanglement wedges with zero cross section, both the reflected entropy and the mutual information indicate a departure of $\rho_{XY}$ from product form. This motivates a study of $S_R(X:Y) - I(X:Y)$, called the Markov gap \cite{hayden2021markov}, which was found to be a marker of genuine tripartite entanglement \cite{akers2020entanglement, zou2021universal}. On the other hand, recent work showed that $S_R(X:Y)$ is not monotonous under partial trace~\cite{Hayden2023notAcorrelationMeasure}.

This section computes a family of reflected R{\'e}nyi entropies $S^{(p,q)}_R (ER:N)$, which are defined below. However, because the reflected R{\'e}nyis do not have a good $q \to 1^+$ limit, they do not afford a replica trick computation of the reflected entropy $S_R (ER:N)$. We explain the technical reason for this, which is similar to but distinct from analogous technical issues reported in \cite{akers2022reflected}. To the extent that the $q \geq 2$ reflected R{\'e}nyi entropies characterize the reflected entanglement spectrum, they indicate a flatter spectrum in the old black hole regime than do their non-reflected counterparts. 

\subsection{Review and setup}
We start with a brief review. Given a density matrix $\rho_{XY}$, we consider its canonical purification $|\Psi^{(1)}\rangle=|\sqrt{\rho_{XY}}\rangle$. It is a normalized pure state on a duplicated Hilbert space $\mathcal{H}_X\otimes\mathcal{H}_Y\otimes\mathcal{H}_{X^*}\otimes\mathcal{H}_{Y^*}$, from which $YY^*$ can be traced it. This defines a reduced density matrix $\rho_{XX^*}$ whose von Neumann entropy is the reflected entropy: $S_R(X:Y)\equiv-{\rm tr}\,(\rho_{XX^*}\log \rho_{XX^*})$. We primarily study the reflected R{\'e}nyi entropies ${\rm tr}(\rho_{XX^*}^q)$. 

Neither the logarithm nor the square root of a density matrix is easy to compute directly. In this paper, as in many works on this subject, we try to circumvent these problems using the replica trick. We use two replica indices; $p$ is employed to approach $\sqrt{\rho_{XY}}$ while $q$ is used to approach $\log \rho_{XY}$ by analytic continuation. 

\begin{figure}[t]
\centering
\includegraphics[width=0.98\textwidth]{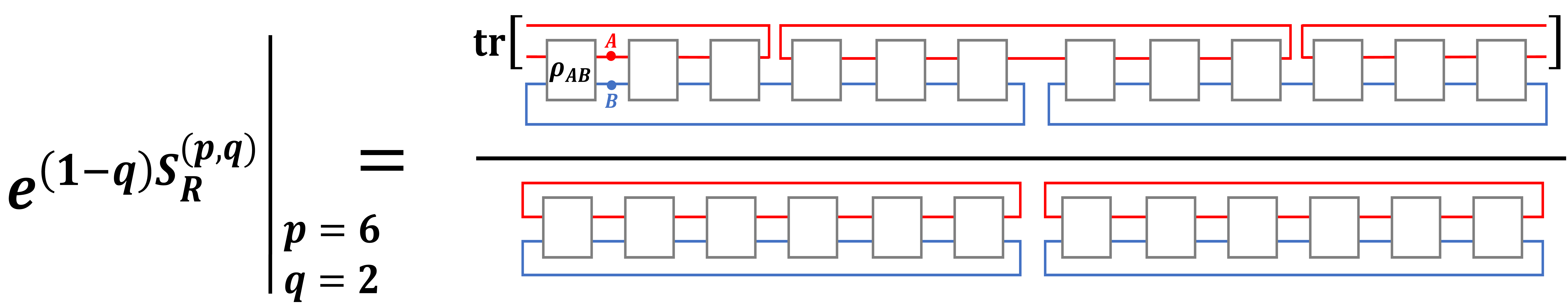}\\
\medskip\medskip\medskip\medskip
\includegraphics[width=0.78\textwidth]{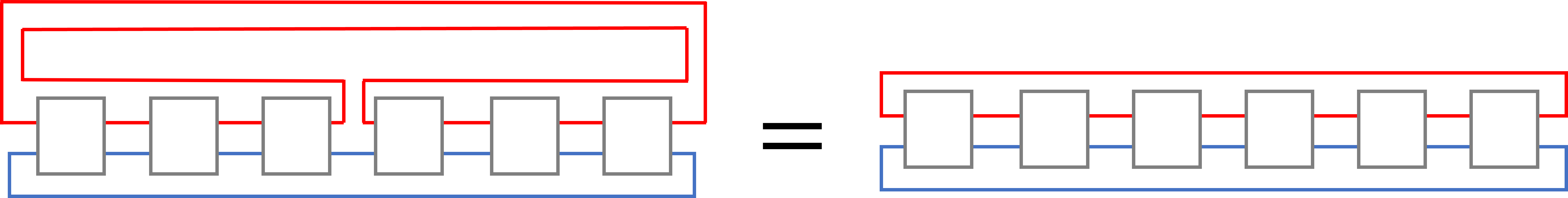}
\caption{Upper panel: An example calculation of $e^{(1-q)S_R^{(p,q)}(A:B)}$, here shown for $p=6$ and $q=2$. The same calculation at $q=1$ must give $e^{(1-q) S_R^{(p,q)}} = 1$; it does so because of the diagrammatic relation in the lower panel. }
\label{fig:reflected}
\end{figure}

As a R{\'e}nyi generalization of the square root, following \cite{dutta2021canonical} we define the normalized state:
\begin{equation}
|\Psi^{(p)}\rangle=\left[\text{tr}(\rho^{p}_{XY})\right]^{-1/2}|\rho^{p/2}_{XY}\rangle
\end{equation}
Formally, substituting $p=1$ returns the canonical purification $|\sqrt{\rho_{XY}}\rangle$, which is used in the definition of the reflected entropy. With an eye to computing the reflected mutual information---specifically, to recover $\log \rho_{XX^*}$ by analytic continuation---we also introduce the replica index $q$, which is more familiar. Altogether, we obtain a family of reflected $(p,q)$-R{\'e}nyi entropies $S_R^{(p,q)}(X:Y)$ given by:
\begin{equation}
e^{(1-q)S_R^{(p,q)}(X:Y)}\equiv\text{tr}\left[\left(\text{tr}_{YY^*}\left[|\Psi^{(p)}\rangle\langle\Psi^{(p)}|\right]\right)^q\right]
=\frac{\text{tr}\left[\left(\text{tr}_{YY^*}\left[|\rho_{XY}^{p/2}\rangle\langle\rho_{XY}^{p/2}|\right]\right)^q\right]}
{\left[\text{tr}(\rho^p_{XY})\right]^q}
\label{defsrpq}
\end{equation}
An example of this expression is represented diagrammatically in Figure~\ref{fig:reflected}.

We denote the numerator on the RHS of (\ref{defsrpq}) as $Z^{(p,q)}(\rho_{XY})$. On the other hand, the denominator is related in a simple way to the $p$-th R\'enyi entropy $S^{(p)}(\rho_{XY})$. Observe that when $q=1$ expression~(\ref{defsrpq}) is tautologically equal to 1. The RHS matches this tautology because $Z^{(p,q=1)}(\rho_{XY}) = \text{tr}(\rho^p_{XY})$, as shown in Figure~\ref{fig:reflected}. This section computes $S_R^{(p,q)}(X:Y)$ explicitly for $q \geq 2$. 

Altogether, the reflected R\'enyi entropy is explicitly given as:
\begin{equation}
S_R^{(p,q)}(X:Y)=\frac{1}{1-q}\log Z^{(p,q)}(\rho_{XY})-\frac{q(1-p)}{1-q}S^{(p)}(\rho_{XY})
\label{eq:renyireflected}
\end{equation}
Because of the tautology in (\ref{defsrpq}) at $q=1$, $\log Z^{(p,q)}(\rho_{XY}) + q (p-1) S^{(p)}(\rho_{XY})$ should have a factor of $1-q$ if these reflected R{\'e}nyi entropies are to have a smooth $q \to 1^+$ limit. This is a precondition for a replica trick calculation of $S_R(X:Y)$. 

\subsection{Old black holes}
\label{sec:reflectedold}
We are interested in the reflected entropy $S_R^{(p,q)}(ER:N)$ between the aggregate radiation $ER$ and the diary catalogue $N$. 
Notice that in the old black hole regime $B'$---the purifier of $ERN$---is maximally mixed, so $S^{(p)}(\rho_{ERN}) = b' = e-c$. Therefore, our task reduces to calculating $Z^{(p,q)}(\rho_{ERN})$.

The structure of the calculation is similar to Section~\ref{sec:renyi}. However, we now average the action of $U \in U(EN)$ on $pq$ copies of the Hilbert space. The resulting permutations $\sigma, \sigma' \in S_{pq}$ close into loops after being paired with new permutations:
\begin{align}
\pi_{B'}&=\iota={\rm identity}\\
\pi_{E}=\pi_{R}&=\xi_p= \Big(1\, 2 \ldots p\Big) \Big((p+1)\, (p+2) \ldots 2p\Big) \ldots \Big((pq-p+1) \ldots (pq)\Big)
\label{eq:xi_1}\\
\pi_{N}&=\xi'_p=\Big((p/2+1)\, (p/2+2) \dots (3p/2)\Big) \ldots \Big((pq-p/2+1) \ldots (pq)\, 1\, 2 \ldots (p/2)\Big)
\label{eq:xi_2}
\end{align}
The right hand side denotes the action of $\xi_p$ and $\xi'_p$ on copies of the $EN$ Hilbert space in cycle notation. In words, $\xi_p$ permutes cyclically $p$ copies of $EN$, and does so in $q$ separate cycles. This traces $(ER)(ER)^*$ out of $p$ copies of $\rho_{ERN}$, and does so $q$ times over. The action of $\xi'_p$ is the same as $\xi_p$ but the cycles are staggered by $p/2$. In each of its $q$ cycles, $\xi'_p$ prepares the bra and the ket part of $\text{tr}_{(ER)(ER)^*}|\rho_{ERN}^{p/2}\rangle\langle\rho_{ERN}^{p/2}|$ and multiplies them. 

After contracting all indices into loops, we obtain the following expression:
\begin{equation}
Z^{(p,q)} = 
(EN)^{-pq} \sum_{\sigma, \sigma' \in S_{pq}} {\rm Wg}(\sigma \sigma'^{-1}) 
B'^{l(\iota, \sigma)} R^{l(\sigma, \xi_p)} E^{l(\xi_p, \sigma')} N^{l(\sigma', \xi'_p)}
\label{eq:Z(p,q)full}
\end{equation}
In principle, this calculates the exact $U(EN)$-average of $Z^{(p,q)}(\rho_{ERN})$ for all integer $q$ and all even integer $p$. 

Similar to the discussion in Section~\ref{sec:renyi}, we would like to simplify (\ref{eq:Z(p,q)full}) in the limit of physical interest where $E,B'\gg R,N$. 
After substituting $B' = EN/R$, using equation~(\ref{wgscaling}) and simplifying terms, we obtain this expression for $Z^{(p,q)}$:
\begin{equation}
\sum_{\sigma,\sigma'\in S_{pq}}{\rm wg}(\sigma\sigma'^{-1})
E^{-\big(d(\iota,\sigma)+d(\sigma,\sigma')+d(\sigma',\xi_p)\big)}
N^{-\big(d(\xi_p,\sigma)+d(\sigma,\sigma')+d(\sigma',\xi'_p)\big)}
(R/N)^{d(\iota, \sigma) - d(\sigma, \xi_p)}
\end{equation}
Terms leading in $E$ and $B'$ minimize:
\begin{equation}
C_{e}\equiv d(\iota,\sigma)+d(\sigma,\sigma')+d(\sigma',\xi_p) \geq d(\iota, \xi_p) = q (p-1).
\label{defce}
\end{equation}

The $\sigma,\sigma'$ that saturate this lower bound live on geodesics from $\iota$ and $\xi_p$ with respect to the Cayley distance. Moreover, $\sigma'$ is closer to $\xi_p$ than $\sigma$, a fact we denote $\sigma' \geq \sigma$. We further observe that $\xi_p$ is a $q$-fold tensor product of generators of $\mathbb{Z}_p$. By the corollary highlighted in Section~\ref{sec:old}, this implies that all geodesics from $\iota$ to $\xi_p$ are obtained by tensoring geodesics from the identity to $\xi \in S_p$ (a single $p$-cycle), and all permutations visited on the way are non-crossing. Putting these facts together and truncating terms subleading in $E$, we obtain:
\begin{equation}
Z^{(p,q)}=E^{-q(p-1)}
\sum_{\substack{\sigma, \sigma' \in ({\rm NC}_p)^{\otimes q}\\ \sigma \leq \sigma'}}
{\rm wg}(\sigma\sigma'^{-1})
N^{-\big(d(\xi_p,\sigma)+d(\sigma,\sigma')+d(\sigma',\xi'_p)\big)}
(R/N)^{d(\iota, \sigma) - d(\sigma, \xi_p)}
\label{refloldfull}
\end{equation}

\paragraph{Terms dominant when diary size is larger than other scales} 
To make progress, we assume $N \gg R/N$, similar to the analysis in Section~\ref{sec:old}. Then the leading term in (\ref{refloldfull}) minimizes
\begin{equation}
C_{n}\equiv d(\xi_p',\sigma')+d(\sigma',\sigma)+d(\sigma,\xi_p) \geq d(\xi'_p , \xi_p) = 2(q-1)
\label{defcn}
\end{equation}
subject to $C_e = q(p-1)$. Minimizing~(\ref{defcn}) calls for a geodesic from $\xi'_p$ to $\xi_p$, which visits $\sigma'$ then $\sigma$ then $\xi_p$ in that order, that is $\sigma \geq \sigma'$. Therefore, simultaneously saturating (\ref{defce}) and (\ref{defcn}) is only possible if $\sigma = \sigma'$ and occurs for those $\sigma$'s, which lie on the intersection of geodesics $\xi_p$-$\iota$ and $\xi_p$-$\xi'_p$. We now explain that this intersection is only $\xi_p$ itself. Later on the same argument will also help us to find subleading corrections. 

The Cayley distance $d(\xi_p, \xi'_p)$ is the minimal number of transpositions $\tau_i$ such that:
\begin{equation}
\tau_{d(\xi_p, \xi'_p)}\, \tau_{d(\xi_p, \xi'_p)-1} \ldots \tau_2\, \tau_1 = \xi'_p (\xi_p)^{-1}
\label{xixi}
\end{equation}
Each set of transpositions $\tau_i$, which realizes the definition of $d(\xi_p, \xi'_p)$ in this way, also defines a geodesic trajectory from $\xi_p$ to $\xi'_p$ given by:
\begin{equation}
\xi_p \,\to\, \tau_1 \xi_p \,\to\, \tau_2 \tau_1 \xi_p \,\to \ldots \to\, 
\tau_{d(\xi_p, \xi'_p)}\, \tau_{d(\xi_p, \xi'_p)-1} \ldots \tau_2\, \tau_1 \xi_p = \xi'_p
\label{trajectoryd}
\end{equation}
Therefore, characterizing geodesics from $\xi_p$ to $\xi'_p$ boils down to finding transpositions $\tau_i$, which satisfy~(\ref{xixi}). Repeating the same reasoning for $d(\xi_p, \iota)$, we conclude that geodesics from $\xi_p$ to $\iota$ are in one-to-one correspondence with sets of transpositions $\kappa_i$ that obey: 
\begin{equation}
\kappa_{d(\xi_p, \iota)}\, \kappa_{d(\xi_p, \iota)-1} \ldots \kappa_2\, \kappa_1 = (\xi_p)^{-1}
\label{xiiota}
\end{equation}
We claim that no transposition can appear both in a solution of (\ref{xixi}) and in a solution of (\ref{xiiota}). This means that geodesics from $\xi_p$ to $\xi'_p$ and from $\xi_p$ to $\iota$ have nothing in common except the initial permutation $\xi_p$. 

To substantiate the claim, we give a prescription for solving~(\ref{xixi}) and (\ref{xiiota}). Formula~$d(\xi_p, \xi'_p) = pq - l(\xi_p, \xi'_p)$ gives another interpretation of Cayley distance: it counts transpositions, which are necessary to break up cycles of $\xi'_p (\xi_p)^{-1}$ into one-cycles. Adopting equation~(\ref{trajectoryd}) to this cycle-based perspective, we say that a set of transpositions $\tau_i$ realizes the Cayley distance $d(\xi_p, \xi'_p)$ if and only if the sequence of permutations
\begin{equation}
\xi'_p (\xi_p)^{-1} \,\to\, \xi'_p (\xi_p)^{-1} \tau_1 \,\to\, \xi'_p (\xi_p)^{-1} \tau_1 \tau_2 \,\to \ldots
\to\, \xi'_p (\xi_p)^{-1} \tau_1 \tau_2 \ldots \tau_{d(\xi_p, \xi'_p)} = \iota
\end{equation}
has an always-growing number of cycles and terminates at $\iota$. In particular, a transposition $\tau_i$ is part of a solution of (\ref{xixi}) if and only if it breaks a cycle of $\xi'_p (\xi_p)^{-1}$. Similarly, a transposition $\kappa_i$ is part of a solution of (\ref{xiiota}) if and only if it breaks a cycle of $(\xi_p)^{-1}$. 

A transposition breaks a cycle if and only if it switches two elements present in that cycle. This fact reduces our claim about the disjointness of the $\xi_p$-$\xi'_p$ and $\xi_p$-$\iota$ geodesics to a simple statement: If two elements live in the same cycle in $\xi'_p (\xi_p)^{-1}$ then they live in distinct cycles in $(\xi_p)^{-1}$ (and therefore also in distinct cycles in $\xi_p$). We easily confirm this by comparing $\xi_p$ in equation~(\ref{eq:xi_1}) to:
\begin{align}
\xi'_p (\xi_p)^{-1} = &
\Big(1\quad (p+1) \quad (2p+1)\, \ldots\, ((q-1)p+1) \Big) \nonumber \\
& \Big( (q-1/2)p+1 \quad (q-3/2)p+1\, \ldots\, (3p/2+1) \quad (p/2+1) \Big)
\end{align}
This concludes the argument that at $N \gg R/N$ equation~(\ref{refloldfull}) is dominated by the unique term $\sigma = \sigma' = \xi_p$. 

\paragraph{First subleading terms} This argument allows us to easily identify the first subleading terms in (\ref{refloldfull}). Such terms are still leading in $E$, which requires that we have $\iota \leq \sigma \leq \sigma' \leq \xi_p$ on a geodesic from $\iota$ to $\xi_p$. We are looking for ways to move $\sigma, \sigma'$ away from $\xi_p$ and pay the least price for it. The least price is $N^{-2}$; that $N^{-1}$ is impossible follows from a simple argument based on the parity of the permutations.

Thus, our goal is to count $\sigma, \sigma'$ such that $\iota \leq \sigma \leq \sigma' \leq \xi_p$
and 
\begin{equation}
C_{n}\equiv d(\xi_p',\sigma')+d(\sigma',\sigma)+d(\sigma,\xi_p) = 2q.
\label{subleadingn}
\end{equation}
The condition from the $\iota$-$\xi_p$ geodesic leaves two possibilities:
\begin{itemize}
\item Move $\sigma$ away from $\xi_p$ by one transposition but leave $\sigma' = \xi_p$. Relative to $\sigma' = \sigma = \xi_p$, this automatically adds $+2$ to $C_n$ because $d(\sigma',\sigma)=d(\sigma,\xi_p)=1$. If $\sigma$ is closer to $\iota$ than $\xi_p$, it must break up a cycle of $\xi_p$. There are $q p (p-1)/2$ such transpositions because $\xi_p$ has $q$ cycles, each of which can be broken in $p(p-1)/2$ distinct ways. Such terms are dressed with a factor $-(R/N)^{q(p-1)-2}$; the sign comes from $\rm{wg(transposition)} = -1$. 
\item Move $\sigma = \sigma'$ away from $\xi_p$. For $q \geq 2$, there are $2^q - 1$ different ways of doing so. 
\end{itemize}

To understand the counting of subleading terms with $\sigma = \sigma' \neq \xi_p$, consider the combination of two geodesic segments $\xi_p$-$\xi_{p/2}$ and $\xi_{p/2}$-$\xi'_p$, where:
\begin{equation}
\xi_{p/2} = \Big(1\, 2 \ldots p/2\Big) \Big((p/2+1)\, (p/2+2) \ldots p\Big) \ldots \Big((pq-p/2+1) \ldots (pq)\Big)
\end{equation}
The cycle structure of $\xi_{p/2}$ is a simultaneous refinement of $\xi_p$ and of $\xi'_p$. Going from $\xi_p$ to $\xi'_p$ via $\xi_{p/2}$ is also a `geodesic' in the sense of being a `local minimum.' But it is not a \emph{global} minimum of the distance function because its length is $2q$. 

Going from $\xi_p$ to $\xi_{p/2}$ keeps us on a geodesic trajectory toward $\iota$ because it only involves breaking up cycles of $\xi_p$. We conclude that any $\sigma$, which satisfies $\xi_{p/2} \leq \sigma \leq \xi_p$, occasions an $N$-subleading term in (\ref{refloldfull}). There are $2^q - 1$ such terms because $\xi_p$ has $q$ cycles, which can be already broken or still unbroken. The $-1$ excludes the term where all remain unbroken because that is the leading term $\sigma = \sigma' = \xi_p$. While we are at it, we can easily account for powers of $R/N$. There are $q! / k! (q-k)!$ terms at distance $d(\sigma, \xi_p) = k$, which come dressed with $(R/N)^{q(p-1)-2k}$. 

\paragraph{Reflected entropy in the old black hole regime}
Putting together all the leading and subleading terms, we find:
\begin{align}
Z^{(p,q \geq 2)} 
& = E^{-q(p-1)} N^{-2(q-1)} (N/R)^{-q(p-1)}
- \frac{qp(p-1)}{2} E^{-q(p-1)} N^{-2q} (N/R)^{-q(p-1)+2} \nonumber \\
& \quad + E^{-q(p-1)} N^{-2q} \sum_{k=1}^q \binom{q}{k} (N/R)^{-q(p-1)+2k} 
+ \mathcal{O}\left(E^{-q(p-1)} N^{-2q-2}\right)
\nonumber \\
& = B'^{-q(p-1)} N^{-2(q-1)} \left[ 
1 - \frac{qp(p-1)N^2}{2R^2} + \frac{1}{N^2} \left( \left( 1 + \frac{N^2}{R^2} \right)^q - 1 \right)
+ \ldots  
\right]
\label{zpqold}
\end{align}
In the last line we used $B' = EN/R$ and performed the sum over $k$.

Using equation~(\ref{eq:renyireflected}) and $S^{(p)}(\rho_{ERN}) = b'$ we obtain:
\begin{equation}
S_R^{(p,q \geq 2)}(ER:N) = 2n + e^{-2n} \left[ \frac{qp(p-1)}{2(q-1)}e^{-2c} 
- \frac{1}{q-1} \bigg( \big( 1 + e^{-2c}\big)^q - 1 \bigg) \right] + \mathcal{O}(e^{-4n})
\label{srpqoldfull}
\end{equation}
If the overhead $R/N = e^{r-n} = e^c$ is of order unity, this expression does not simplify further. But if the overhead is large (though still much smaller than the diary), that is if $n \gg c \gg 1$, then we have:
\begin{equation}
S_R^{(p,q \geq 2)}(ER:N) = 2n + e^{-2r}\, \frac{q(p+1)(p-2)}{2(q-1)} + \mathcal{O}(e^{-2r-2c})
\label{srpqoldsimple}
\end{equation}

We see that the subleading corrections do not have a good $q \to 1$ limit.\footnote{This difficulty is somewhat similar but distinct from the one encountered in \cite{akers2022reflected}, where the reflected entropy of a single random tensor and of random tensor network (RTN) states was considered. There, the difficulty appears already at the level of describing the dominant saddle point whereas here the difficulty concerns subleading terms.} To salvage the replica trick, we could try sending simultaneously $p,q \to 1$. In that case, the correction terms that originate from $\sigma \neq \sigma' = \xi_p$ may be alright because they come with a factor $(p-1)/(q-1)$. But terms with $\sigma = \sigma' \neq \xi_p$ continue to cause a problem. This is understandable because the subleading character of such terms is established by trajectories, which approach $\xi_{p/2}$ by splitting $p$-cycles into halves. As stated, these operations do not have a limit $\lim_{p \to 1}$.

We may speculate that such terms should simply be discarded in a replica calculation of $I_R(ER:N) = \lim_{p,q \to 1} S_R^{(p,q)}(ER:N)$ because they capture a peculiarity of even $p$. Some circumstantial evidence for this prescription is that including the problematic terms replaces a factor of $p-1$ with a factor of $p-2$. Still, the status of the calculation at the level of subleading terms is uncertain. If we trust the discarding prescription then the subleading corrections come at relative order $e^{-2r}$. This would indicate a much flatter reflected entanglement spectrum than the ordinary entanglement spectrum, which shows features at order $e^{-2c}$; viz. equation~(\ref{iqernold}). 

\paragraph{How does the $q=1$ computation work?} Result~(\ref{srpqoldsimple}) does not have a $q \to 1$ limit yet we know $Z^{(p,q=1)}(\rho_{ERN}) = \text{tr}(\rho^p_{ERN}) = B'^{(1-p)}$. What corrects $Z^{(p,q)}$ in equation~(\ref{zpqold}) to recover this $q=1$ identity? 

To understand this, return to condition~(\ref{subleadingn}), which selects the subleading terms. Below it, we identified two classes of subleading terms: those with $\sigma \neq \sigma' = \xi_p$ and those with $\sigma = \sigma' \neq \xi_p$.  When $q=1$, the count of the former terms is correct at $p(p-1)/2$; this means that \emph{every} transposition $\tau$ yields one subleading term $\sigma = \tau \xi_p$. But the latter class of terms has a different counting at $q=1$ and at $q\geq 2$. 

Working at $q \geq 2$, we counted terms $\sigma = \sigma' \neq \xi_p$ by counting permutations on the segment $\xi_p$-$\xi_{p/2}$. Such permutations are special because they live on the common part of geodesics $\xi_p$-$\iota$ and $\xi_p$-$\xi'_p$. When $q=1$, however, the permutations $\xi_p$ and $\xi'_p$ are one and the same permutation! In particular, $\xi_{p/2}$ plays no role in the analysis. Instead, for \emph{every} transposition $\tau$, setting $\sigma = \sigma' = \tau \xi_p$ satisfies equation~(\ref{subleadingn}). Consequently, at $q=1$, the putative subleading terms from families $\sigma \neq \sigma' = \xi_p$ and $\sigma = \sigma' \neq \xi_p$ come in pairs, which are distinguished only by $\rm{wg(}\iota {\rm{)}} = 1$ and $\rm{wg(transposition)} = -1$. In other words, they cancel out. In fact, the cancelation extends to all terms with $\sigma \neq \xi$; see Appendix~\ref{sec:weingarten}. This leaves out $Z^{(p,q=1)} = B'^{(1-p)} $, which is of course mandated by the tautology $Z^{(p,q=1)}(\rho_{ERN}) = \text{tr}(\rho^p_{B'})$ and by $\rho_{B'} = 1/B'$.

\subsection{Young black holes}
\label{reflectedyoung}
We now turn to young black holes, which are not maximally entangled with their early radiation. As in Section~\ref{sec:young}, we model this feature by initializing the black hole and radiation in a random state on $EB$ to be averaged over. The loop counting is the same as in Section~\ref{sec:reflectedold} except that the factor $E^{l(\xi_p, \sigma')}$ splits into $E^{l(\xi_p, \tau)} B^{l(\tau,\sigma')}$, like it did in Section~\ref{sec:young}. After this substitution, we find for $Z^{(p,q)}(\rho_{ERN})$:
\begin{equation}
[N^{pq}(BE)^{pq}f(BE)]^{-1}
\sum_{\tau,\sigma,\sigma'\in  S_{pq}}{\rm Wg}(\sigma\sigma'^{-1})
B'^{l(\iota, \sigma)}R^{l(\sigma,\xi_p)}N^{l(\xi'_p, \sigma')}B^{l(\sigma', \tau)}E^{l(\tau, \xi_p)}
\label{zpqyoungreflected}
\end{equation}

We assume that the size of the black hole dominates over all other independent parameters, including the early radiation ($B \gg E$). Then, after substituting $B' = BN/R$, we find that the overall exponent of $B$ is the negative of:
\begin{equation}
C_b=d(\iota,\sigma)+d(\sigma,\sigma')+d(\sigma',\tau) \geq d(\iota, \tau)
\end{equation}
Minimizing this---that is, maximizing the exponent of $B$---sets $\sigma = \sigma' = \tau = \iota$, like it did in Section~\ref{sec:young}. If we truncate~(\ref{zpqyoungreflected}) to leading order, we obtain
\begin{equation}
\log Z^{(p,q)}(\rho_{ERN}) = q(1-p) (e+r+n) + \ldots = q(1-p) S^{(q)}(\rho_{ERN}) + \ldots
\end{equation}
where the second equality compares the calculation to our earlier result~(\ref{eq:rho_ERN_young}). Using equation~(\ref{eq:renyireflected}) we see that leading order terms cancel out in $S_R^{(p,q)}(ER:N)$. Thus, the reflected entropy of the diary and radiation is exponentially small in black hole size, as is expected for a young black hole.

To compute this exponentially small quantity we need the subleading terms in (\ref{zpqyoungreflected}). They come in the same four varieties, which we encountered in Section~\ref{sec:young}: (i) the first subleading term in $[f(BE)]^{-1}$ and the cases where (ii) $\tau$ alone or (iii) $\sigma'=\tau$ or (iv) $\sigma=\sigma'=\tau$ is a single transposition. However, in contrast to the calculation of ${\rm tr}\,(\rho_{ERN})^q$ in Section~\ref{sec:young}, the exponents and combinatorial coefficients within each class are no longer uniform across all transpositions. 

It is useful to highlight the difference between the present calculation and expression~(\ref{trrhonfullyoung}) in Section~\ref{sec:young}. Recall that the latter involves permutation $\xi \in S_q$, which is a cycle of maximal length and which therefore gets broken by the action of any transposition.\footnote{This resembles the present calculation at $q=1$, where $\xi_p = \xi'_p$. From here on, we assume $q \geq 2$.}  Equation~(\ref{zpqyoungreflected}) has a similar structure but $\xi$ is replaced by two distinct permutations $\xi_p$ and $\xi'_p$, none of which is a cycle of maximal length. Consequently, transpositions can either break (B) or glue (G) cycles of $\xi_p$ and $\xi'_p$. This divides transpositions in $S_{pq}$ into four classes, which we denote BB, BG, GB, GG. (The first letter refers to the action on the cycles of $\xi_p$ while the second letter refers to the action on $\xi'_p$.) 

We can organize first subleading terms in $Z^{(p,q)}(\rho_{ERN})$ in a 4x4 array, whose rows correspond to the four classes of transpositions and columns correspond to the cases (i-iv). That is, $Z^{(p,q)}(\rho_{ERN}) = (ERN)^{q(1-p)}$ plus
\begin{equation}
\begin{array}{rlll} 
(ERN)^{q(1-p)}B^{-1}
\bigg(~~\phantom{+}C_{{\rm BB}} \Big(-E^{-1} &~ + E &~ - E & + ER^2\Big) \\
\phantom{\bigg(}+C_{{\rm BG}} \Big(-E^{-1} &~ + E &~ -EN^{-2} & + ER^2 N^{-2}\Big) \\
\phantom{\bigg(}+C_{{\rm GB}} \Big(-E^{-1} &~ +E^{-1} &~ -E^{-1} &+ E^{-1}\Big) \\
\phantom{\bigg(}+C_{{\rm GG}}\Big(-E^{-1} &~ +E^{-1}&~ -E^{-1}N^{-2}& + E^{-1}N^{-2}\Big)\bigg)
\end{array}
\label{4x4}
\end{equation}
plus further terms of order $(ERN)^{q(1-p)} B^{-2}$. Notice that the BB row mimics the structure of correction terms in (\ref{eq:rho_ERN_young}) because the latter describe the effect of cycle-breaking transpositions. 

The largest term in the array is $ER^2$. It dominates over all others unless $N$ is of order unity. (We need not stipulate $R \sim 1$ separately because $R > N$.) After dropping all other terms, our last task is to count the transpositions in the BB class, denoted $C_{\rm BB}$. 

For even $p$, a transposition $(x \, y)$ breaks a cycle of $\xi_p$ and a cycle of $\xi_p$ if and only if $x$ and $y$ are in the same cycle of $\xi_{p/2}$. (In terms of partitions, $\xi_{p/2}$ is the coarsest common refinement of $\xi_p$ and $\xi_{p/2}$.) Thus, we are counting ways to break one of the $2q$ cycles of $\xi_{p/2}$. Since a cycle of length $p/2$ can be broken in $(1/2)(p/2) (p/2-1)$ ways, we find:
\begin{equation}
C_{\rm BB} = qp(p-2)/4
\end{equation}

Putting all the results together, we find:
\begin{equation}
\log Z^{(p,q)}(\rho_{ERN}) = q(1-p) (e+2n + c) + \frac{qp (p-2)}{4} e^{- (b-e)+ 2(n+c)} + \mathcal{O}(e^{-(b-e)+2c})
\end{equation}
We remind the reader that $e^c = R/N$ quantifies the radiation overhead used in siphoning Alice's diary out of Hawking radiation. The correction terms are exponentially small because $B = e^b$ dominates over all other scales in the problem. Using equations~(\ref{eq:renyireflected}) and (\ref{ernq}), we finally arrive at the reflected $(p,q)$-R{\'e}nyi entropies: 
\begin{equation}
S_R^{(p,q)}(X:Y) 
= \frac{qp^2}{4(q-1)} e^{-(b-e) + 2(n+c)}
\label{youngreflectedfinal}
\end{equation}

Once again, this expression does not have a good $q \to 1^+$ limit. If we set $q=1$ then $\xi_p = \xi'_p$ and \emph{all} transpositions in $S_{pq} = S_p$ are of the BB type. In that case, calculation~(\ref{4x4}) becomes identical to (\ref{eq:rho_ERN_young}), as is mandated by the tautology $Z^{(p,q=1)}(\rho_{ERN}) = \text{tr}(\rho^p_{ERN})$. 

Observe that the failure of (\ref{youngreflectedfinal}) to have a good $q \to 1^+$ limit has the same origin as in the old black hole calculation. It arises from the fact that $\xi_p \neq \xi'_p$ when $q \geq 2$ but $\xi_p = \xi'_p$ when $q=1$, which selects different sets of permutations to contribute at the same order. Unlike here, in the old black hole in equation~(\ref{srpqoldsimple}) this phenomenon appears at a subleading order because the $\mathcal{O}(1)$ reflected R{\'e}nyi entropy is nonvanishing. However, the first exponentially suppressed order in both (young and old) calculations shows the same qualitative feature.  

\section{Discussion}
\label{sec:discussion}

This paper inspects the recovery of information from unitarily evaporating black holes. In keeping with prior works and the spirit of Hayden and Preskill's original argument, we model black hole evaporation as a random unitary process to be averaged over. In addition, for black holes before Page time, we also average over the initial state of the black hole-early radiation subsystem. For both old and young black holes, we have computed R{\'e}nyi entropies of the diary reference $N$, the radiation $ER$, and the remaining black hole $B'$. These quantities encode the spectra of the respective density matrices, which we extract in Appendix~\ref{sec:spectra}.

Naturally, our computations confirm the main conclusion of \cite{Hayden2007mirror}---that is, after Page time black holes return the information dropped into them. We have also gleaned several additional facts, which give a fuller picture of information recovery:
\begin{itemize}
\item We have computed the first subleading corrections of all the R{\'e}nyi entropies, which characterize the information recovery process. In old black holes they are exponentially suppressed in radiation overhead (intercepted radiation minus diary size); viz. equations~(\ref{iqernold}-\ref{iqnbold}). In young black holes; the suppressing parameter is $b-e -2r$; viz. equations~(\ref{iqernyoung}-\ref{iqnbyoung}). This is the `Page gap' $b-e$ (the entropy gap that separates the black hole from becoming old) minus $2r$, i.e. double the radiation collected after the diary was dropped. If we reset $e \to e+r$ and $b \to b-r$ and close the Page gap then $b-e = 2r$. Thus, the suppression parameter is the excess Page gap after taking the radiation into account. 
\item In young black holes, we have identified a range of intercepted radiation sizes, over which the mutual informations involving $N$ (the diary reference), $ER$ (the collected radiation) and $B'$ (the remaining black hole) transition between their minimal and maximal values; see equation~(\ref{rrange}). 
This is the range when the information about the diary is split between the radiation $ER$ and the black hole $B'$ in a commensurate proportion. It is summarized in Figure~\ref{fig:entropies}. 
\item The beginning of this range is when the radiation is just enough to close the `Page gap' without accounting for the diary: $e+r_{\rm init} = b-r_{\rm init}$. The duration of the transitional range is set by the size of the diary $n$. The end of the transitional period is when radiation $r_{\rm final} = (b-e)/2 + n$ is collected, at which point the total radiation system $ER$ is no longer approximately maximally mixed. The beginning and end of the transitional range are identified by subleading terms in $I(ER:N)$ becoming order unity.
\item If the diary had been part of the black hole from the very beginning (resetting $b \to b+n$) then the radiation system $ER$ would cease to be maximally mixed when $e + r = b + n - r$, that is when $r = (b-e+n)/2$. By tossing a diary into a young black hole after it has emitted radiation $E$, we reset this threshold time to $r_{\rm final} = (b-e+2n)/2$. Thus, depositing new information in a young black hole effectively delays its Page time. Studies in humans have shown that continual learning delays the onset of symptoms of old age \cite{humanage}; here we reach the same conclusion regarding black holes. 
\item We have also analyzed the special regime where the `Page gap' $b-e$, which separates a young black hole from being old is not parametrically large. This allows us to relate and contrast computations in young and old black holes. The sharpest contrast is offered by equations~(\ref{tomin-large}), (\ref{btomax}) and (\ref{midagemax}).
\end{itemize}

Our calculations can be easily modified and extended for other purposes. One example of a similar calculation is \cite{akers2022reflected}, where R{\'e}nyi and reflected R{\'e}nyi entropies were calculated in random tensor network (RTN) states. That paper also includes a complete analysis of a random tensor state on three legs. (In that language, our calculations for old black holes characterize random tensor states on four legs, where two legs have much larger bond dimensions.) Regarding future applications, one natural variation on our calculations would be to randomly split the early radiation into two components---one intercepted by Bob and one that he inadvertently lost. Recoverability of information from incomplete early Hawking radiation has been recently discussed in \cite{Bao:2020zdo, withsirui, vijayspaper}; our paper sets up a tractable way to further study this question.

Finally, we have also attempted to compute the reflected mutual information $I_R(ER:N)$ via the replica trick. We computed the reflected R{\'e}nyi entropies $S_R^{(p,q)}(ER:N)$ in equations~(\ref{srpqoldsimple}) and (\ref{youngreflectedfinal}) but they do not admit a replica trick continuation to $q \to 1^+$. We identified a common technical origin of the problem, which afflicts both the old and young black hole calculation: different sets of permutations contribute at the first exponentially suppressed order at $q=1$ and at $q\geq 2$. We have characterized this difference in detail.

If we try to convert the reflected R{\'e}nyi entropies (\ref{srpqoldsimple}, \ref{youngreflectedfinal}) into spectral densities, we get divergent von Neumann entropies; see Appendix~\ref{sec:spectra}. This is a big caveat on using our calculations to characterize the reflected entanglement spectra. If we do so, however, we conclude that nontrivial features of the reflected entanglement spectrum are exponentially suppressed by $2r$ in the old black hole case, and by $(b-e)-2r$ in the young black hole case. For a young black hole, this is the same characteristic scale as in the non-reflected spectrum; see equations~(\ref{iqernyoung}-\ref{iqnbyoung}). For an old black hole, however, our results would indicate a much flatter reflected entanglement spectrum than the non-reflected spectrum. The latter shows features suppressed only by the radiation overhead $c$, as seen in equations~(\ref{iqernold}-\ref{iqnbold}).

\acknowledgments
An early version of the material in this paper was first presented by HT as a final project in class \textit{Quantum Information and Quantum Gravity}, which BC taught (and SS attended) at Tsinghua University in Fall 2022. We thank other students in the class, especially Dachen Zhang, for useful discussions. BC and SS thank the organizers of the workshop `Quantum Information, Quantum Matter and Quantum Gravity' (YITP-T-23-01) held at YITP, Kyoto University (2023), where this work was completed. The work of BC and SS is supported by an NSFC grant number 12042505, a BJNSF grant under the Gao Cengci Rencai Zizhu program, and a Dushi Zhuanxiang Fellowship.

\appendix
\section{Weingarten functions}
\label{sec:weingarten}

In the main text, we frequently use products of matrix elements of unitary matrices, averaged over the unitary group. Such averages are characterized by Weingarten functions \cite{weingartenoriginal, kostenberger}. 

Specifically, we have in mind an integral over the unitary group $U(D)$ with Haar measure $dU$ given by:
\begin{equation}
\int dU\cdot U_{i_1j_1}\cdots U_{i_qj_q}U^\dagger_{j'_1 i'_1}\cdots U^\dagger_{j'_q i'_q} 
= \sum_{\sigma,\tau\in S_q}
\delta_{i_1i'_{\sigma(1)}}\cdots\delta_{i_q i'_{\sigma(q)}}
\delta_{j'_1 j_{\tau^{-1}(1)}}\cdots\delta_{j'_q j_{\tau^{-1}(q)}} {\rm Wg}(\sigma\tau^{-1}) \nonumber
\end{equation}
Here $\sigma,\tau$ are permutations in $S_q$; the coefficients ${\rm Wg}(\sigma \tau^{-1})$ are called Weingarten functions. The Weingarten functions depend on Hilbert space dimension $D$. However, in this paper we always average over $U(EN)$ or $U(BN)$ (when $B \neq E$), so we drop the $D$-dependence from our notation to avoid clutter. 

This paper uses the $D \gg 1$ scaling of the Weingarten functions, which is:
\begin{equation}
{\rm Wg}(\sigma) = 
D^{-n-d(\sigma, \iota)}\, {\rm wg}(\sigma)+\mathcal{O}\big(D^{-n-d(\sigma, \iota)-2}\big)
\label{eq:appendix:Wg's scaling}
\end{equation}
For $\sigma \in S_q$, which is a product of cycles of length $c_i$, the leading order coefficient ${\rm wg}(\sigma)$ is:
\begin{equation}
{\rm wg}(\sigma) = \prod_i \frac{(-1)^{c_i-1}\,(2c_i-2)!}{(c_i-1)!\,c_i!}
\end{equation}
Special cases, which are used in the paper, include ${\rm wg}(\iota) = 1 = - {\rm wg}({\rm transposition})$.

\paragraph{Averaging over pure states} 
The averaged tensor product of $|\psi\rangle \langle \psi|$, which is used in Sections~\ref{sec:young} and \ref{reflectedyoung}, is also normalized by an expression involving Weingarten functions. This fact is best derived diagrammatically: 
\begin{equation}
\centering
\includegraphics[width=0.8\textwidth]{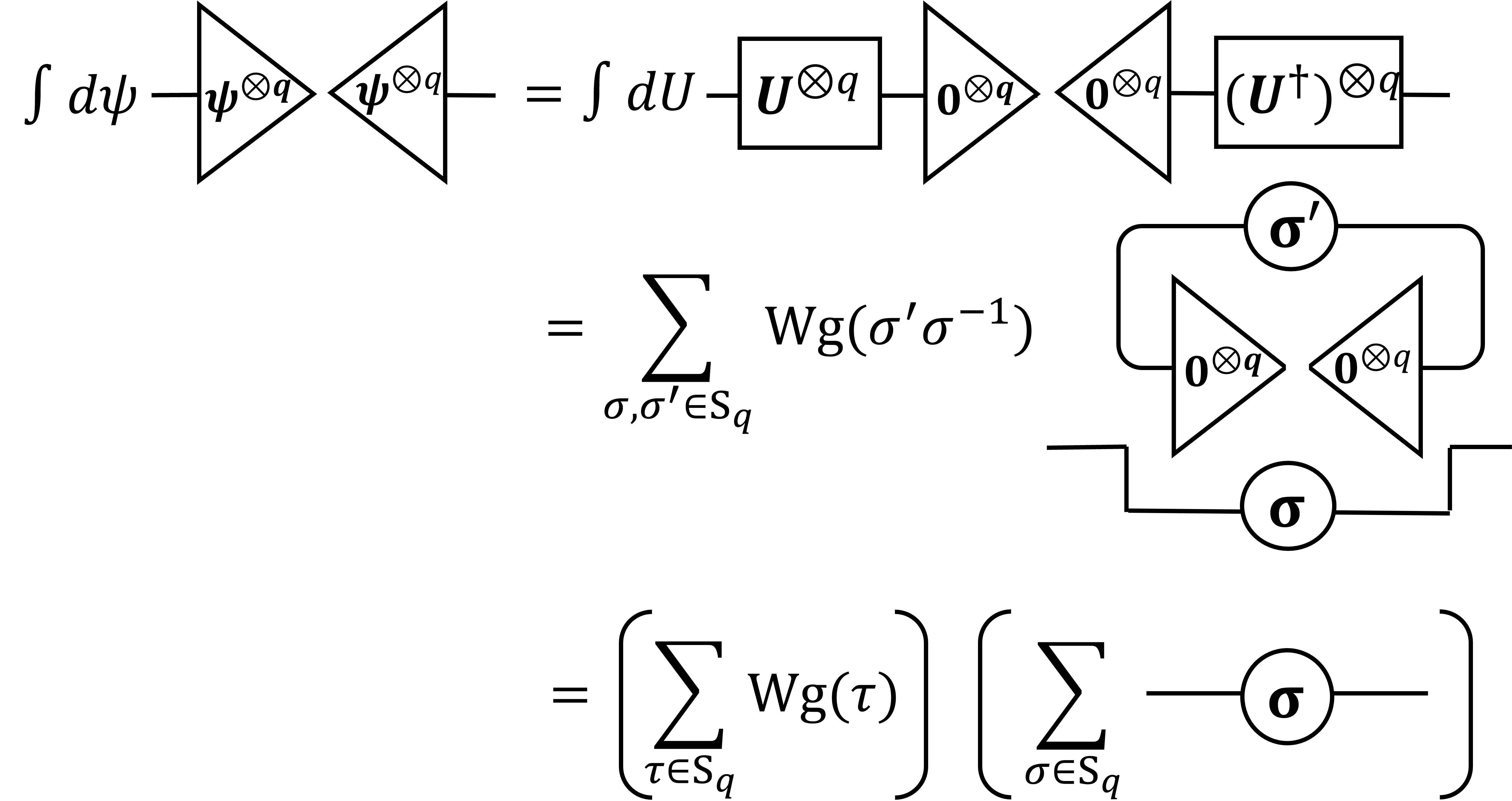}
\label{weingartenf}
\end{equation}
In the main text, we normalize this integral with $\overline{|\psi\rangle\langle\psi|^{\otimes q}} = [D^q f(D)]^{-1} \sum_\sigma \Pi_\sigma$. The normalization factor can be written in various ways; some of them were used in equation~(\ref{deffnorm}) but (\ref{weingartenf}) gives us another neat rewriting \cite{kostenberger, stanley}:
\begin{equation}
D^q f(D) 
= \frac{(D+q-1)!}{(D-1)!} 
= \sum_{\sigma\in S_q}D^{l(\sigma,\iota)} 
= \left(\sum_{\tau \in S_q} {\rm Wg}(\tau)\right)^{-1}
\end{equation}

\paragraph{An exact calculation from Weingarten functions}
Here we calculate the moments ${\rm tr} (\rho_{B'}^q)$ of an old black hole using Weingarten functions. As we observed in Section~\ref{sec:old}, this density operator is identically $\rho_{B'} = 1/B'$ because the unitaries $U$ and $U^\dagger$ cancel pairwise in the diagrammatic expression for $\rho_{B'}$. Therefore, the calculation must give ${\rm tr} (\rho_{B'}^q) = B'^{1-q}$. We present this calculation for several reasons:
\begin{itemize}
\item It enables a comparison with analogous calculations where $\rho_{B'} \neq 1/B'$. For example, equation~(\ref{fullexactapp}) below can be contrasted with equation~(\ref{trrhonfullyoung}), which applies to young black holes.
\item It enables a comparison between the $q=1$ and $q \geq 2$ calculations of quantity $Z^{(p,q)}(\rho_{ERN})$ in Section~\ref{sec:reflectedold}; see equation~(\ref{eq:Z(p,q)full}). Equation~(\ref{fullexactapp}) below is $Z^{(p,q=1)}(\rho_{ERN})$.
\item It showcases property~(\ref{indicator}) of Weingarten functions.  
\end{itemize}
In the formalism of this paper, quantity ${\rm tr} (\rho_{B'}^q)$ for a young black hole is written as:
\begin{equation}
{\rm tr}\,(\rho_{B'})^q = 
{\rm tr}\,(\rho_{ERN})^q = 
(EN)^{-q} \sum_{\sigma, \sigma'\in S_q} {\rm Wg}(\sigma \sigma'^{-1}) 
B'^{l(\iota, \sigma)} R^{l(\sigma, \xi)} N^{l(\sigma', \xi)} E^{l(\sigma', \xi)}
\label{fullexactapp}
\end{equation}
We recall that $D = EN = B'R$. By applying the identity \cite{kostenberger, zinnjustin}
\begin{equation}
\sum_{\sigma' \in S_q} {\rm Wg}(\sigma \sigma'^{-1}) D^{l(\sigma', \chi)} = \delta_{\sigma,\chi}
\label{indicator}
\end{equation}
we obtain:
\begin{align}
{\rm tr}\,(\rho_{B'})^q & = 
(BR)^{-q} \sum_{\sigma \in S_q} B'^{l(\iota, \sigma)} R^{l(\sigma, \xi)} 
\sum_{\sigma'\in S_q} {\rm Wg}(\sigma \sigma'^{-1}) D^{l(\sigma', \xi)} 
\nonumber \\
& = (BR)^{-q} B'^{l(\iota, \xi)} R^{l(\xi, \xi)} = B'^{1-q}
\label{fullexactapp2}
\end{align}

\section{Entanglement spectra}
\label{sec:spectra}

The R{\'e}nyi entropies computed in Section~\ref{sec:renyi} are, up to multiplication by $(1-q)$ and exponentiation, moments of the density matrix ${\rm tr}(\rho^q)$. This data allows us to recover the spectrum of the density matrix, expressed as the density of eigenvalues
\begin{equation}
D(\lambda)\equiv\sum_{i}\delta(\lambda-\lambda_i) 
\end{equation}
where ${\lambda_i}$ are the eigenvalues of $\rho$. To relate $D(\lambda)$ to the moments, use the rewriting of the delta function $\delta(x)=\pi\lim_{\epsilon\rightarrow0}{\rm Im}(\frac{1}{x-i\epsilon})$ to find:
\begin{equation}
D(\lambda) 
= \pi\lim_{\epsilon\rightarrow0}\,{\rm Im} \sum_i \frac{1}{\lambda - \lambda_i - i\epsilon} 
= \pi\lim_{\epsilon\rightarrow0}\,{\rm Im}\, {\rm tr}\, (\lambda - i\epsilon - \rho)^{-1}
\end{equation}
It is convenient to define the Green's function
\begin{equation}
G(\lambda) = {\rm tr}(\lambda-\rho)^{-1} 
= \lambda^{-1}\sum_{q=0}^{\infty}\lambda^{-q}\,\,{\rm tr}(\rho^q),
\label{eq:G(lambda)}
\end{equation}
which satisfies:
\begin{equation}
D(\lambda)=\pi\lim_{\epsilon\rightarrow0}{\rm Im}\,G(\lambda-i\epsilon)
\label{eq:D(lambda)}
\end{equation}

\paragraph{Spectrum of density matrix}
For reader's convenience, we copy the non-trivial moments of density matrices, which were computed in Section~\ref{sec:renyi}:
\begin{align}
{\rm tr}(\rho_{ER}^q)&=(EN^2/R)^{1-q}\left(1+\frac{q(q-1)}{2}\exp(-2c)\right)&& \text{(old black hole)}\\
{\rm tr}(\rho_{ER}^q)&=(ER)^{1-q}\left(1+\frac{q(q-1)}{2}\exp(-b+e+2c)\right)&& \text{(young black hole)}\\
{\rm tr}(\rho_{ERN}^q)&=(ERN)^{1-q}\left(1+\frac{q(q-1)}{2}\exp(-b+e+2n+2c)\right)&& \text{(young black hole)}
\end{align}
Each of them can be parametrized in the form:
\begin{equation}
{\rm tr}(\rho^q)=\beta^{1-q}\left(1+\frac{q(q-1)}{2}\alpha\right)
\end{equation}
with $\beta\leq D_{\rho}\equiv{\rm tr}(1)$ and $\alpha>0$ a known, exponentially small parameter.

Using equations~(\ref{eq:G(lambda)}) and (\ref{eq:D(lambda)}), we find:
\begin{align}
&G(\lambda)=\frac{D_{\rho}-\beta}{\lambda}+\frac{\beta}{\lambda-\beta^{-1}}+\frac{\alpha\beta^{-1}}{(\lambda-\beta^{-1})^3}\\
&D(\lambda)=(D_{\rho}-\beta)\delta(\lambda)+\beta\delta(\lambda-\beta^{-1})+\frac{\alpha}{2\beta}\delta''(\lambda-\beta^{-1})
\end{align}
Direct integration confirms that this density of eigenvalues returns the correct values of:
\begin{equation}
{\rm tr}(\rho^q) = \int d\lambda \, D(\lambda)\, \lambda^q
\end{equation}

\paragraph{Spectrum of reflected density matrix}
In equations~(\ref{srpqoldsimple}) and (\ref{youngreflectedfinal}), we calculated nontrivial moments of the reflected density matrix $\rho\equiv{\rm tr}_{NN^*}\left[|\Psi^{(p)}_{ERN}\rangle\langle\Psi^{(p)}_{ERN}|\right]$ of the radiation of an old and young black hole. They take the general form
\begin{equation}
{\rm tr}(\rho^q)= \beta^{1-q} \big( 1 - q \alpha\, \delta_{q \neq 1}  \big)
\label{eq:refls}
\end{equation}
with parameters:
\begin{align}
\beta & = N^2 & & {\rm and} & \alpha & = \frac{(p+1)(p-2)}{2} e^{-2r} & & \textrm{(old black hole)} \\
\beta & = 1     & & {\rm and} & \alpha & = \frac{p^2}{4} e^{-(b-e)+2r} & & \textrm{(young black hole)}
\end{align}
Then the Green's function and spectral density are given by:
\begin{align}
&G(\lambda)=\frac{D_{\rho}-\beta}{\lambda}+\frac{\beta}{\lambda-\beta^{-1}}
-\alpha\left(\frac{1}{(\lambda-\beta^{-1})^2}-\frac{1}{\lambda^2}\right)\\
&D(\lambda)=(D_{\rho}-\beta)\delta(\lambda)+\beta \delta(\lambda-\beta^{-1})+\alpha \left(\delta'(\lambda-\beta^{-1})-\delta'(\lambda)\right)
\label{reflectedspectrum}
\end{align}
Here $D_{\rho}\equiv{\rm tr}(1)=(ER)^2$ is the dimension of $\rho$. 

Integrating $\int d\lambda\, D(\lambda) \lambda^q$ gives back (\ref{eq:refls}), which is a self-consistency check on (\ref{reflectedspectrum}). However, the density of eigenvalues implied by (\ref{eq:refls}) gives a divergent reflected entropy:
\begin{equation}
-{\rm tr}(\rho\log\rho)
=- \int d\lambda \, D(\lambda)\, \lambda \log \lambda 
= (1-\alpha) \log \beta -\alpha \int d\lambda\, \delta(\lambda) \log\lambda
\end{equation}
We observe that this divergence comes from the $\lambda^{-2}$ term appearing in $G(\lambda)$, which in turn originates from the discontinuity of ${\rm tr}(\rho^q)$ between $q=1$ and $q\geq2$.

\end{document}